\documentclass[journal,hidelinks]{IEEEtran}

\usepackage{graphicx,url}
\usepackage[english]{babel}
\addto\captionsenglish{}
\usepackage[utf8]{inputenc}
\usepackage{amsmath}
\usepackage{amssymb}
\usepackage{lipsum}
\usepackage{mathtools}
\usepackage{cuted}
\usepackage{cite}
\usepackage{graphicx}
\usepackage{balance}
\usepackage{tikz}
\usepackage{pgfplots}
\usepackage{tikzscale} 
\pgfplotsset{compat=newest} 
\usepackage{scalefnt}
\usepackage{orcidlink}
\usepackage{dblfloatfix}
\usepackage[acronym,shortcuts]{glossaries}
\usepackage{algorithm, algpseudocode}
\usepackage{algcompatible}
\usepackage{bm}

\newacronym{RMSE}{RMSE}{root mean square error}
\newacronym{MMSE}{MMSE}{minimum mean square error}
\newacronym{MF}{MF}{matched filter}
\newacronym{RPE}{RPE}{radar parameter estimation}
\newacronym{OTFS}{OTFS}{orthogonal time frequency space}
\newacronym{AFDM}{AFDM}{affine frequency division multiplexing}
\newacronym{MIMO}{MIMO}{multiple-input multiple-output}
\newacronym{SISO}{SISO}{single-input single-output}
\newacronym{ISAC}{ISAC}{integrated sensing and communications}
\newacronym{3D}{3D}{three-dimensional}
\newacronym{2D}{2D}{two-dimensional}
\newacronym{1D}{1D}{one-dimensional}
\newacronym{RX}{RX}{receiver}
\newacronym{TX}{TX}{transmitter}
\newacronym{BF}{BF}{beamforming}
\newacronym{mmWave}{mmWave}{millimeter-wave}
\newacronym{SotA}{SotA}{state-of-the-art}
\newacronym{ULA}{ULA}{uniform linear array}
\newacronym{QAM}{QAM}{quadrature amplitude modulation}
\newacronym{ISFFT}{ISFFT}{inverse symplectic finite Fourier transform}
\newacronym{SFFT}{SFFT}{symplectic finite Fourier transform}
\newacronym{AWGN}{AWGN}{additive white Gaussian noise}
\newacronym{OFDM}{OFDM}{orthogonal frequency division multiplexing}
\newacronym{OCDM}{OCDM}{orthogonal chirp division multiplexing}
\newacronym{BS}{BS}{base station}
\newacronym{UE}{UE}{user equipment}
\newacronym{DFT}{DFT}{discrete Fourier transform}
\newacronym{IDFT}{IDFT}{inverse discrete Fourier transform}
\newacronym{IFFT}{IFFT}{inverse fast Fourier transform}
\newacronym{TD}{TD}{time-domain}
\newacronym{wlg}{wlg}{without loss of generality}
\newacronym{CP}{CP}{cyclic prefix}
\newacronym{DAFT}{DAFT}{discrete affine Fourier transform}
\newacronym{DAF}{DAF}{discrete affine Fourier}
\newacronym{IDAFT}{IDAFT}{inverse discrete affine Fourier transform}
\newacronym{CPP}{CPP}{\textit{chirp-periodic} prefix}
\newacronym{IDZT}{IDZT}{inverse discrete Zak transform}
\newacronym{DZT}{DZT}{discrete Zak transform}
\newacronym{ICI}{ICI}{inter-carrier interference}
\newacronym{BER}{BER}{bit error rate}
\newacronym{DoF}{DoF}{degrees-of-freedom}
\newacronym{FD}{FD}{full-duplex}
\newacronym{SIMO}{SIMO}{single-input multiple-output}
\newacronym{MISO}{MISO}{multiple-input single-output}
\newacronym{AoD}{AoD}{angle-of-departure}
\newacronym{AoA}{AoA}{angle-of-arrival}
\newacronym{RF}{RF}{radio frequency}
\newacronym{SIM}{SIM}{stacked intelligent metasurfaces}
\newacronym{FPGA}{FPGA}{field programmable gate array}
\newacronym{UPA}{UPA}{uniform planar array}
\newacronym{CC}{CC}{communication-centric}
\newacronym{I/O}{I/O}{input-output}
\newacronym{iid}{i.i.d.}{independent and identically distributed}
\newacronym{IoT}{IoT}{internet of things}
\newacronym{V2X}{V2X}{vehicle-to-everything}
\newacronym{NTN}{NTN}{non-terrestrial network}
\newacronym{LEO}{LEO}{low-earth orbit}
\newacronym{THz}{THz}{terahertz}
\newacronym{EM}{EM}{expectation maximization}
\newacronym{RIS}{RIS}{reconfigurable intelligent surface}
\newacronym{DoA}{DoA}{direction-of-arrival}
\newacronym{DD}{DD}{doubly-dispersive}
\newacronym{ODDM}{ODDM}{orthogonal delay-Doppler division multiplexing}
\newacronym{LoS}{LoS}{line-of-sight}
\newacronym{NLoS}{NLoS}{non-line-of-sight}
\newacronym{6G}{6G}{sixth generation}
\newacronym{MPDD}{MPDD}{metasurfaces-parametrized DD}
\newacronym{GaBP}{GaBP}{Gaussian belief propagation}
\newacronym{MSE}{MSE}{mean-squared-error}
\newacronym{sIC}{soft IC}{soft interference cancellation}
\newacronym{soft RG}{soft RG}{soft replica generation}
\newacronym{BG}{BG}{belief generation}
\newacronym{SGA}{SGA}{scalar Gaussian approximation}
\newacronym{CLT}{CLT}{central limit theorem}
\newacronym{PDF}{PDF}{probability density function}
\newacronym{QPSK}{QPSK}{quadrature phase-shift keying}
\newacronym{OQAM}{OQAM}{offset quadrature amplitude modulation}
\newacronym{LMMSE}{LMMSE}{linear minimum mean square error}
\newacronym{SNR}{SNR}{signal-to-noise ratio}
\newacronym{OOBE}{OOBE}{out-of-band emission}
\newacronym{PAPR}{PAPR}{peak-to-average power ratio}
\newacronym{AFBM}{AFBM}{affine filter bank modulation}
\newacronym{FBMC}{FBMC}{filter bank multicarrier modulation}
\newacronym{PPN}{PPN}{polyphase network}
\newacronym{SIR}{SIR}{signal-to-interference ratio}
\newacronym{AF}{AF}{ambiguity function}
\newacronym{PDA}{PDA}{probabilistic data association}
\newacronym{SBL}{SBL}{sparse Bayesian learning}
\newacronym{VGA}{VGA}{vector Gaussian approximation}
\newacronym{KL}{KL}{Kullback-Leibler}
\newacronym{GAMP}{GAMP}{generalized approximate message passing}
\newacronym{EP}{EP}{expectation propagation}
\newacronym{5G}{5G}{fifth generation}
\newacronym{4G}{4G}{fourth generation}

\newcommand\scalemath[2]{\scalebox{#1}{\mbox{\ensuremath{\displaystyle #2}}}}

\newcommand{\trans}[0]{^{\mathsf{T}}}

\newcommand{\herm}[0]{^{\mathsf{H}}}

\begin{document}

\title{Affine Filter Bank Modulation (AFBM): A Novel\\ 6G ISAC Waveform with Low PAPR and OOBE}
%

\author{Kuranage Roche Rayan Ranasinghe\textsuperscript{\orcidlink{0000-0002-6834-8877}}, \IEEEmembership{Graduate Student Member,~IEEE,}
Henrique L. Senger\textsuperscript{\orcidlink{0009-0004-1586-8168}}, \\
Gustavo P. Gonçalves\textsuperscript{\orcidlink{0009-0000-8260-4390}}, 
Hyeon Seok Rou\textsuperscript{\orcidlink{0000-0003-3483-7629}}, \IEEEmembership{Member,~IEEE,}
Bruno S. Chang\textsuperscript{\orcidlink{0000-0003-0232-7640}}, \IEEEmembership{Member,~IEEE,} \\
Giuseppe Thadeu Freitas de Abreu\textsuperscript{\orcidlink{0000-0002-5018-8174}}, \IEEEmembership{Senior Member,~IEEE,} 
and Didier Le Ruyet\textsuperscript{\orcidlink{0000-0002-9673-2075}}, \IEEEmembership{Senior Member,~IEEE}

\thanks{K.~R.~R.~Ranasinghe, H~S.~Rou and G.~T.~F.~de~Abreu are with the School of Computer Science and Engineering, Constructor University, Campus Ring 1, 28759 Bremen, Germany (emails: [kranasinghe, hrou, gabreu]@constructor.university).}
\thanks{H.~L.~Senger, G.~P.~Gonçalves and B.~S.~Chang are with the CPGEI/Electronics Department, Federal University of Technology - Paraná, Curitiba, Brazil (emails: [hsenger, gustavog.1999]@alunos.utfpr.edu.br, bschang@utfpr.edu.br).}
\thanks{D.~Le~Ruyet is with the LAETITIA/CEDRIC/Conservatoire National des Arts et Métiers - Paris, France (email: didier.le$\_$ruyet@cnam.fr).}
\thanks{Preliminary parts of this work have been accepted for presentation at the 2025 IEEE International Workshop on Computational Advances in Multi-Sensor Adaptive Processing (CAMSAP) \cite{ranasinghe2025lowcomplexityreceiverdesignaffine}.}
\vspace{-5ex}
}

\maketitle

\begin{abstract} 
We propose the \ac{AFBM} waveform for enhanced \ac{ISAC} in \ac{6G}, designed by drawing on concepts from classical \ac{FBMC} theory and recent advances in chirp-domain waveforms, particularly \ac{AFDM}. 
Specifically, \ac{AFBM} exhibits several desirable properties, with emphasis on its remarkably low \ac{PAPR} and reduced \ac{OOBE} when benchmarked against the conventional \ac{AFDM} waveform under \ac{DD} channel conditions. 
In the communications setting, reliable symbol detection is achieved using a tailored low-complexity \ac{GaBP}-based algorithm, while in the sensing setting, a range and velocity estimation approach is developed that integrates an \ac{EM}-assisted \ac{PDA} framework to accurately identify surrounding targets. 
The highlighted performance and benefits of \ac{AFBM} are validated through analytical and numerical evaluations, including conventional metrics such as \ac{AF}, \ac{BER}, and \ac{RMSE}, consolidating its position as a promising waveform for next-generation wireless systems.
\end{abstract}


\IEEEpeerreviewmaketitle

\glsresetall

\section{Introduction}

\IEEEPARstart{T}{he} next generation of wireless communication systems is envisioned to deliver exceptionally high data rates by exploiting higher frequency bands \cite{JiangCOMMST2024,KangOJCOMS2024}, while at the same time supporting a wide range of advanced applications such as high-mobility communications \cite{Jingxian2016,XueWCOM2025,ZhangTWC2025,MuWCL2025,WangTWC2025}, computing functionalities \cite{Ranasinghe_ICNC2025,RanasingheTWC2025ICC,ZhangTCOM2025,YanTCOM2025,LiuTWC2025}, and \ac{ISAC} \cite{Zhiqing_IoT2023, NuriaISAC2024, RanasingheTWC2025,ZhangVTM2025,CorreasAESM2025,WeiIEEENET2024}. 
These high-mobility and high-frequency scenarios inherently give rise to \ac{DD} channels, which induce severe inter-carrier interference and significantly degrade the performance of \ac{OFDM} \cite{Svante2007,SahinTCOM2025}, the waveform adopted in \ac{4G} and \ac{5G} systems.

In response to these challenges, a variety of waveform designs have been investigated \cite{Rou_SPM_2024}, with \ac{OTFS} and \ac{AFDM} standing out as two of the most promising alternatives. 
The \ac{OTFS} waveform, introduced in \cite{HadaniWCNC2017} as a \ac{2D} modulation scheme that directly embeds information in the delay-Doppler domain, has attracted significant attention due to its inherent \ac{ISAC}-enabling properties \cite{GaudioTWC2020}. 
However, despite these advantages, \ac{OTFS} suffers from higher modulation complexity than \ac{OFDM}, fails to achieve the optimal diversity order in \ac{DD} channels \cite{SurabhiTWC2019}, and lacks backward compatibility with existing hardware infrastructure \cite{rou2025affine}. 
As an evolution of \ac{OTFS}, the \ac{ODDM} waveform \cite{TongTCOM2024} achieves reduced modulation complexity and lower \ac{PAPR}, but this comes at the expense of reduced diversity. 
On the other hand, the \ac{AFDM} waveform \cite{bemani2023,TaoJSAC2025,LiIEEENET2025} maps data symbols onto chirp-based subcarriers in the \ac{DAF} domain, enabling full diversity in \ac{DD} channels through proper tuning of its chirp parameters. 
Nevertheless, \ac{AFDM} inherits the high \ac{PAPR} and poor spectral containment of \ac{OFDM}, leading to inefficient power amplification and significant \ac{OOBE}.

Several works have sought to alleviate these shortcomings by addressing both \ac{PAPR} and \ac{OOBE}. 
For instance, in \cite{tao2024daft}, the \ac{DFT} operation in spread-\ac{OFDM} (DFT-s-OFDMA) was replaced with a \ac{DAFT}, giving rise to the DAFT-s-AFDMA scheme, which improves \ac{PAPR} performance under certain chirp parameters. 
From a spectral containment perspective, a modified version of \ac{OCDM}, termed c-\ac{OCDM}, was proposed in \cite{omar2020spectrum}, where frequency-domain zero padding was introduced into chirp matrices prior to modulation, resulting in improved spectral localization. 
This idea was later extended to \ac{AFDM} in \cite{savaux2024}, where a modified transceiver structure was proposed that reduces overall computational complexity while also improving spectral containment.

A complementary perspective is provided by \ac{FBMC}, which inherently offers excellent spectral localization due to its per-subcarrier filtering, though it lacks robustness to the \ac{DD} impairments that are crucial in \ac{ISAC}. 
Classical \ac{FBMC} achieves real orthogonality through \ac{OQAM} modulation in order to satisfy the Balian-Low theorem \cite{KorevaarTCOM2016}, while precoded variants \cite{zakaria2012novel, nissel2018pruned, pereira2021novel, pereira2022generalized} approximate complex orthogonality and provide additional benefits such as reduced \ac{PAPR} through pruned \ac{DFT}-precoding. 
Unlike conventional multicarrier waveforms such as \ac{OFDM}, \ac{OTFS}, or \ac{AFDM}, \ac{FBMC} does not require a \ac{CP}, which improves spectral efficiency and reduces energy overhead, while its prototype filters minimize \ac{OOBE}, making it well suited for fragmented spectrum access. 
The offset structure of \ac{OQAM} modulation further enhances robustness to frequency-selective fading and synchronization errors, but complicates \ac{MIMO} processing and pilot design. 
This tradeoff has motivated substantial research on precoded and generalized \ac{FBMC} schemes that attempt to preserve the desirable spectral characteristics of \ac{FBMC} while alleviating its limitations in multi-antenna and multi-user scenarios. 
As an example, the 2D-FFT \ac{FBMC} waveform was proposed in \cite{pereira2023two}, providing improved \ac{OOBE} characteristics, with a follow-up enhancement in \cite{junior2023generalized} that reduced \ac{PAPR} at the cost of increased computational complexity.

Despite these efforts, a clear gap remains in the literature: no \ac{ISAC}-oriented waveform has yet been proposed that simultaneously achieves robustness against \ac{DD} impairments while also ensuring both inherently low \ac{PAPR} and reduced \ac{OOBE}. 
To address this gap, we propose a novel waveform, termed \ac{AFBM} and first introduced in \cite{senger2025affinefilterbankmodulation}, which is designed to overcome the limitations of existing approaches. 
In particular, \ac{AFBM} integrates a pruned \ac{DAFT} precoding stage with a filter-bank structure into the \ac{AFDM} framework, thereby combining the advantages of these methods. 
As a result, \ac{AFBM} achieves \acp{PAPR} comparable to DAFT-s-AFDM, \ac{OOBE} suppression similar to that of \ac{FBMC}, and quasi-orthogonality in \ac{DD} channels.

Nevertheless, the transceiver design of \ac{AFBM} for both communication and sensing remains largely unexplored, since existing \ac{AFDM}-based approaches cannot be directly applied due to the added complexity introduced by chirp-based subcarriers and filter-bank processing. 
In addition, the domain-specific power distributions of \ac{AFBM} must be carefully considered when designing appropriate receiver structures. 
To this end, we further propose a low-complexity communications receiver based on the \ac{GaBP} framework, which enables symbol detection using only element-wise scalar operations, and a sensing receiver based on the \ac{PDA} framework enhanced with \ac{EM}, which is capable of accurately estimating scatterer ranges and velocities from pilot preambles. 

The remainder of this paper is organized as follows. 
Section~\ref{secSysModel} introduces the system model and describes the modulation procedure of the \ac{AFBM} waveform. 
Section~\ref{sec:AFBM_channel_characteristics} presents the channel analysis of \ac{AFBM} and provides a mathematical justification for the suitable domain in the design of \ac{ISAC} receivers. 
Section~\ref{sec:PAPR_OOBE} evaluates the \ac{PAPR}, \ac{OOBE}, and \ac{AF} performance of \ac{AFBM} in comparison with \ac{SotA} alternatives. 
Section~\ref{secReceivercomm} details the low-complexity communication receiver based on the \ac{GaBP} framework, while Section~\ref{secReceiverISAC} introduces the sensing receiver based on the \ac{PDA} framework. 
Finally, Section~\ref{secConclusion} concludes the paper and discusses directions for future work. 

The main contributions of this paper are summarized by:
\begin{itemize}
\item The novel \ac{AFBM} waveform is proposed and its fundamental properties, including \ac{PAPR}, \ac{OOBE}, \ac{AF}, are analyzed. 
\item A low-complexity communication receiver based on the \ac{GaBP} framework is developed to efficiently estimate data symbols from the unique filtered time-domain channel matrix of \ac{AFBM}.
\item A sensing receiver based on the \ac{PDA} framework, enhanced with \ac{EM}, is designed to estimate target ranges and velocities in a bistatic setting using the pilot preamble and the filtered time-domain channel matrix of \ac{AFBM}.
\end{itemize}

\section{System Model}
\label{secSysModel}

\begin{figure}[H]
\vspace{-2ex}
\centering
\includegraphics[width=\columnwidth]{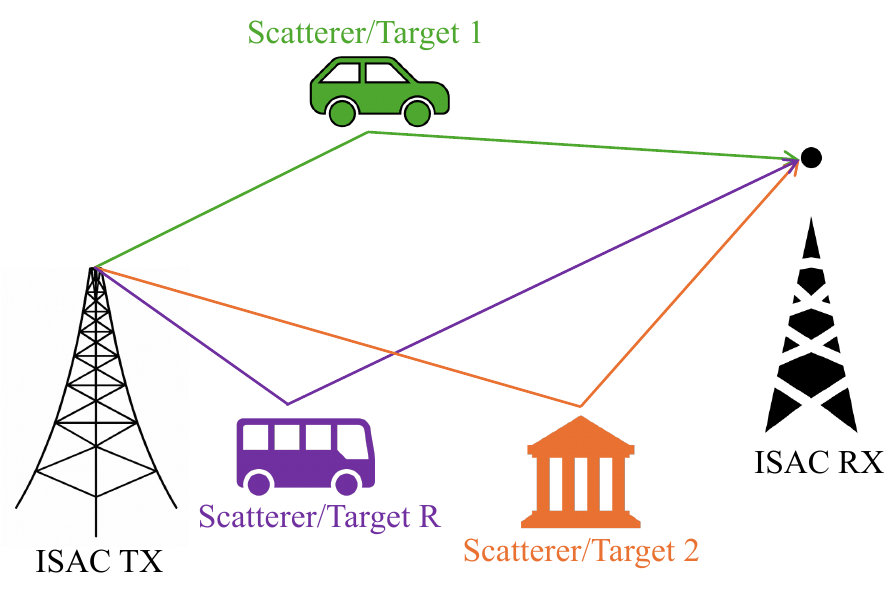}
\vspace{-5ex}
\caption{Illustration of an \ac{AFBM}-\ac{ISAC} scenario, where the \ac{ISAC}-\ac{RX} either i) attempts to decode the data symbols $\mathbf{x}$ under the knowledge of the channel matrix $\mathbf{H}$, or ii) attempts to estimate the target delays and Doppler shifts $(\tau_r, \nu_r), \forall r$ using either a reference signal (pilots) or a pre-existing fronthaul.}
\label{fig:AFBMmod}
\end{figure}

A point-to-point communication system is considered with a single-antenna \ac{ISAC}-\ac{TX} and an \ac{ISAC}-\ac{RX} operating over a \ac{DD} channel with $P$ target scatterers, where the proposed \ac{AFBM} waveform is employed, as illustrated in Fig.~\ref{fig:AFBMmod}. 
As shown in the figure, the model encompasses two scenarios: \textit{i)} the \ac{ISAC}-\ac{RX} estimates the transmitted data $\mathbf{x}$ given knowledge of the channel matrix $\mathbf{H}$, or \textit{ii)} the \ac{ISAC}-\ac{RX} estimates the target parameters $(\tau_p, \nu_p), \forall p$, using either a reference signal (pilots) or a pre-existing fronthaul link.

\subsection{Transmit Signal Model}
\label{subsec:transmit_signal_model}

Let $L$ denote the number of subcarriers and $K$ the number of time indices in a filter bank-based multicarrier system with near complex orthogonality. 
To avoid inter-filter interference, half of the $L$ subcarriers are reserved as guard bands, and transmission is carried out at twice the rate of conventional \ac{OFDM}/\ac{AFDM} systems, namely, every $L/2$ samples (see \cite{Ranasinghe_ICNC2025_oversampling} for a detailed discussion on the oversampling tradeoff). 
The system is organized into blocks of $K$ symbols, each with a duration of $T/2$ seconds. 
Within each symbol, $L$ subcarriers are employed, spaced by $F$ Hz. This results in a grid comprising $L$ points in frequency, spaced by $F$ Hz, and $K$ points in time, spaced by $T/2$ seconds in the time-frequency domain. 
Accordingly, the total bandwidth is given by $B = L F$, and the total transmission interval by $K T/2$.

Let $\mathbf{x} \in \mathcal{D}^{K\frac{L}{2} \times 1}$ denote the vector of transmit symbols mapped onto the defined time-frequency resources, where $\mathcal{D}$ represents a modulation alphabet of size $|\mathcal{D}|$ (e.g., \ac{QAM} constellations). 

\begin{figure*}[t!]
\centering
\includegraphics[width=\textwidth]{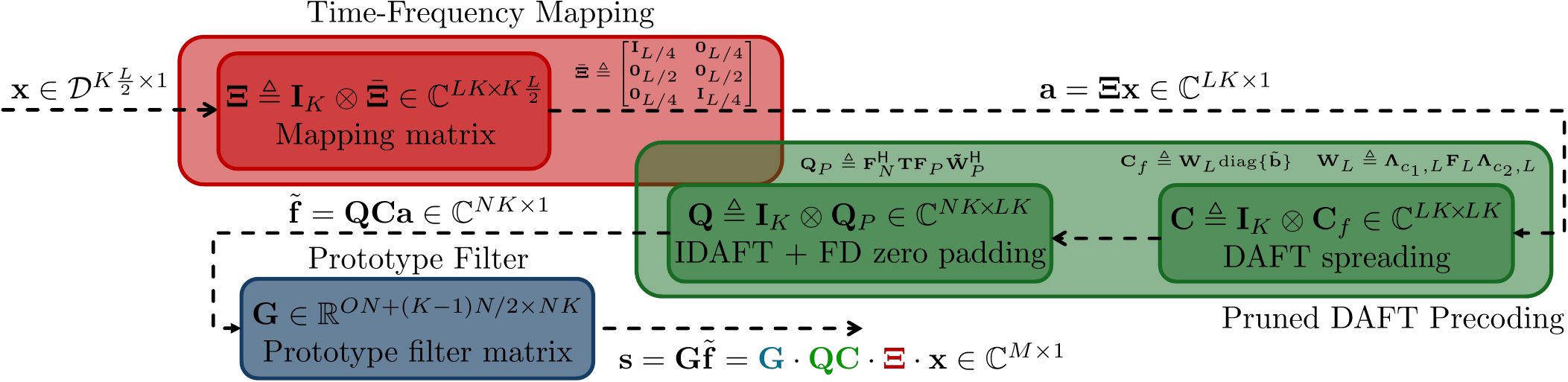}
\vspace{-3ex}
\caption{Visualization of the \ac{AFBM} modulation procedure.}
\label{fig:AFBMmod_scehamtic}
\vspace{-1ex}
\end{figure*}

The symbols in $\mathbf{x}$ are arranged in the first and last $L/4$ positions of a matrix $\mathbf{A} \in \mathbb{C}^{L \times K}$ according to a pre-established transmission strategy, so as to avoid interference from the filter bank. The mapping is expressed as
\begin{equation}
\label{eq:positions}
\mathbf{a} \triangleq \mathrm{vec}(\mathbf{A}) = \bm{\Xi} \mathbf{x} \in \mathbb{C}^{LK \times 1},
\end{equation}
where $\mathrm{vec}(\cdot)$ denotes the column-wise vectorization operation and $\bm{\Xi} \in \mathbb{C}^{LK \times K\frac{L}{2}}$ is defined as
\begin{equation}
\label{eq:Xi}
\bm{\Xi} \triangleq \mathbf{I}_K \otimes \bar{\bm{\Xi}},
\end{equation}
with $\bar{\bm{\Xi}} \in \mathbb{C}^{L \times \frac{L}{2}}$ given by
\begin{equation}
\bar{\bm{\Xi}} \triangleq 
\begin{bmatrix}
\mathbf{I}_{L/4} & \mathbf{0}_{L/4} \\
\mathbf{0}_{L/2} & \mathbf{0}_{L/2} \\
\mathbf{0}_{L/4} & \mathbf{I}_{L/4}  
\end{bmatrix},
\end{equation}
where $\mathbf{0}_L$ denotes a zero matrix of size $L$. 

The matrix $\mathbf{A}$ defined in \eqref{eq:positions} is subsequently multiplied by a diagonal matrix in order to restore complex orthogonality, and then transformed into the \ac{DAFT} domain through a pruned \ac{DAFT} precoding operation. 
In the following, the procedure for restoring complex orthogonality through a filter compensation stage is described, after which the pruned \ac{DAFT} precoding operation is detailed.

\subsubsection{Restoration of Complex Orthogonality}
\label{subsubsec:complex_orthogonality}

A filter-bank waveform with well-localized filters and complex orthogonality can be obtained by employing the \ac{DAFT} together with a filter compensation stage that cancels interference. 
To this end, let us define
\begin{equation}
\mathbf{C}_f \triangleq \mathbf{W}_{L} \mathrm{diag}\{\tilde{\mathbf{b}}\},
\label{ferf44}
\end{equation}
where $\mathbf{C}_f \in \mathbb{C}^{L \times L}$ represents the precoding process responsible for restoring complex orthogonality, and $\mathbf{W}_{L} \in \mathbb{C}^{L \times L}$ denotes the $L$-point DAFT matrix, defined as
\begin{equation}
\mathbf{W}_{L} = \mathbf{\Lambda}_{c_1,L}\mathbf{F}_{L}\mathbf{\Lambda}_{c_2,L},
\end{equation}
with
\begin{equation}
\mathbf{\Lambda}_{c_i,L} = \mathrm{diag}[e^{-j2\pi c_i (0)^2}, \dots, e^{-j2\pi c_i (L-1)^2}] \in \mathbb{C}^{L \times L}
\end{equation}
denoting an $L \times L$ diagonal chirp matrix with central digital frequency $c_i$, and where $\mathbf{F}_{L}$ denotes the normalized $L$-point \ac{DFT} matrix.

To preserve complex orthogonality, $\mathbf{C}_f$ must be chosen such that the following condition is satisfied:
\begin{equation}
\mathbf{C}\herm_f\mathbf{Q}_{P}^{H}\mathbf{\widetilde{G}}^T\mathbf{\widetilde{G}}\mathbf{Q}_{P} \mathbf{C}_f \approx \mathbf{U},
\label{eq:complex_orthogonality}
\end{equation}
where $\mathbf{U} \in \mathbb{R}^{L \times L}$ is a diagonal matrix with unit values in the first and last $L/4$ positions and zeros elsewhere. 

Here, $\mathbf{\widetilde{G}}$ denotes the filtering matrix corresponding to the transmission of a single multicarrier symbol, expressed as $\mathbf{\widetilde{G}} = [\mathbf{G}_0; \mathbf{G}_1; \ldots; \mathbf{G}_{2O-1}] \in \mathbb{R}^{ON \times N}$. 
By substituting \eqref{ferf44} into \eqref{eq:complex_orthogonality}, the $\tilde{l}$-th element of $\tilde{\mathbf{b}}$ is obtained as
\begin{equation}
[\mathbf{\tilde{b}}]_{\tilde{l}} = 
\begin{cases} 
\sqrt{\frac{1}{[\mathbf{\tilde{c}}]_{\tilde{l}}}}, & \tilde{l} \in \left[ 0,\ldots,\tfrac{L}{4}-1 \right] \cup \left[ L-\tfrac{L}{4},\ldots,L-1 \right] \\[1ex]
0, & \text{otherwise},
\end{cases}
\end{equation}
with
\begin{equation}
\mathbf{\tilde{c}} \triangleq \mathrm{diag}\{\mathbf{W}\herm_L\mathbf{Q}_{P}^{H}\mathbf{\widetilde{G}}^T\mathbf{\widetilde{G}}\mathbf{Q}_{P}\mathbf{W}_L\}.
\end{equation}

The compensation stage thus comprises a multiplicative factor that cancels the interference in the transmitted symbols introduced by the filter coefficients. 
Since the coefficients are derived from a pre-defined prototype filter \cite{pereira2023two}, they are assumed to be known. 
Correct compensation is guaranteed when the interference is limited to a single coefficient, which is achieved by selecting an overlap factor $O \leq 1.5$ \cite{pereira2022generalized}. 
If the overlap factor exceeds this threshold, off-diagonal interference appears in \eqref{eq:complex_orthogonality}, thereby reducing the achievable \ac{SIR}. 

\subsubsection{Pruned DAFT Precoding}

Based on \eqref{eq:positions} and \eqref{ferf44}, the vectorized form of the DAFT-spread transmit signal $\mathbf{b} \in \mathbb{C}^{LK \times 1}$, whose matrix form is denoted by $\mathbf{B} \in \mathbb{C}^{L \times K}$ before filtering, is expressed as \vspace{-2ex}
\begin{align}
\label{eq:precodede_tx}
\mathbf{b} & \triangleq \mathrm{vec}(\mathbf{B}) = \mathrm{vec}\big(\overbrace{\mathbf{W}_L\mathrm{diag} (\tilde{\mathbf{b}})}^{\mathbf{C}_f \, \in \, \mathbb{C}^{L \times L}}\mathbf{A}\big) \\
&  = \underbrace{\big(\mathbf{I}_K \otimes \mathbf{C}_f \big)}_{\triangleq \mathbf{C} \, \in \, \mathbb{C}^{LK \times LK}} \mathrm{vec}(\mathbf{A}) =  \mathbf{C} \mathbf{a} = \mathbf{C} \bm{\Xi} \mathbf{x}. \nonumber
\end{align}

\vspace{-1ex} \noindent where the pruned \ac{DAFT} $\mathbf{\tilde{W}}_{P} \in  \mathbb{C}^{L \times P}$ is defined as
\begin{equation}
\mathbf{\tilde{W}}_{P} = 
\begin{bmatrix}
\mathbf{I}_L &  \mathbf{0}_{L\times (P-L)}  
\end{bmatrix}
\mathbf{W}_P.
\label{dft_espalhada}
\end{equation}

Afterwards, by applying frequency-domain zero-padding to the IDAFT output, the output matrix $\mathbf{Q}_{P}$ for a given block is obtained as
\begin{equation}
\mathbf{Q}_{P} = \mathbf{F}_{N}^{H}\mathbf{T}\mathbf{F}_P \mathbf{\tilde{W}}\herm_{P},
\label{eq:Q_P}
\end{equation}
where
\begin{equation}
\mathbf{T}\triangleq 
\begin{bmatrix}
\mathbf{I}_{P/2} & \mathbf{0}_{P/2} \\
\mathbf{0}_{(N-P) \times P/2} & \mathbf{0}_{(N-P) \times P/2} \\
\mathbf{0}_{P/2} & \mathbf{I}_{P/2}  
\end{bmatrix},
\end{equation}
is an $N \times P$ matrix satisfying $\mathbf{T}^{T}\mathbf{T} = \mathbf{I}_P$. 

Considering the transmission of $K$ blocks, the block matrix $\mathbf{Q} \in  \mathbb{C}^{NK\times LK}$ is expressed as
\begin{equation}
\mathbf{Q} =  \mathbf{I}_{K} \otimes \mathbf{Q}_{P}, 
\end{equation}
where $\otimes$ denotes the Kronecker product, mapping $\mathbf{Q}_{P}$ into the correct time positions.

\subsubsection{Prototype Filter}

The transmitted data are obtained by convolving the precoded signal with the prototype filter impulse response through a Toeplitz filter matrix. 
Let $\mathbf{G}_p \in  \mathbb{R}^{N/2 \times N/2}$ denote the diagonal matrix of filter coefficients, i.e.,
\begin{equation}
\mathbf{G}_p = \mathrm{diag}(\mathbf{g}_p), \;\;\; p = 0,1,2,\ldots,2O-1
\end{equation}
where
\begin{equation}
\mathbf{g}_p = [g[pN/2], g[pN/2+1], \ldots , g[pN/2+N/2-1]]
\end{equation}
and $\mathbf{g}$ represents the prototype filter of length $ON$, with $O$ denoting the overlap factor. 

Accordingly, the block Toeplitz filter matrix $\mathbf{G} \in  \mathbb{R}^{ON + (K-1)N/2 \times NK}$ is defined as
\begin{equation}
\mathbf{G} = 
\scalemath{0.8}{\begin{bmatrix}
\mathbf{G}_0  & \mathbf{0} & \mathbf{0} & \mathbf{0} & \ldots  & \mathbf{0} \\
\mathbf{0} & \mathbf{G}_1  & \mathbf{G}_0  & \mathbf{0} & \ldots  & \mathbf{0} \\
\mathbf{G}_2  & \mathbf{0} & \mathbf{0} & \mathbf{G}_1  &  \ldots  & \mathbf{0} \\
\mathbf{0} & \mathbf{G}_3  & \mathbf{G}_2 & \mathbf{0} & \ldots  & \mathbf{0} \\
\vdots &  \mathbf{0} &  \mathbf{0} & \mathbf{G}_3  &  \ldots  & \mathbf{0} \\
\vdots &  \vdots &  \vdots &  \vdots &  \ddots  &   \vdots \\
\mathbf{G}_{2O-4} & \vdots &  \vdots &  \vdots & \ddots  &  \mathbf{0} \\
\mathbf{0} & \mathbf{G}_{2O-3} & \mathbf{G}_{2O-4} &  \vdots & \ddots & \mathbf{G}_1 \\
\mathbf{G}_{2O-2} &  \mathbf{0} &  \mathbf{0} &  \mathbf{G}_{2O-3} & \ddots &  \mathbf{0} \\
\mathbf{0} & \mathbf{G}_{2O-1} & \mathbf{G}_{2O-2} & \mathbf{0} & \ddots &  \mathbf{G}_3 \\
\vdots &  \mathbf{0} & \mathbf{0} & \mathbf{G}_{2O-1} & \ddots & \vdots \\
\vdots &  \vdots & \ddots &   \vdots &  \ddots &  \mathbf{0} \\
\mathbf{0} &   \mathbf{0} &  \ldots & \mathbf{0} &  \ddots & \mathbf{G}_{2O-1} 
\end{bmatrix}}.
\label{fhddg}
\end{equation}

\subsubsection{Effective Transmit Signal}

In all, the complete \ac{AFBM} transmit signal in the \ac{TD} can be expressed in terms of the compensation matrix $\mathbf{C}_f$ in \eqref{ferf44}, the modified \ac{IDAFT} matrix $\mathbf{Q}_P$ in \eqref{eq:Q_P}, and the filter matrix $\mathbf{G}$ in \eqref{fhddg}, by exploiting Kronecker product identities as
\begin{align}
\label{eq:td_tx_signal}
\mathbf{s} & = \mathbf{G} \mathbf{Q} \mathbf{C} \mathbf{a}  
= \mathbf{G} \big(\mathbf{I}_{K} \otimes \mathbf{Q}_{P}\big) \cdot \big(\mathbf{I}_K \otimes \mathbf{C}_f \big) \mathbf{a} \in \mathbb{C}^{M \times 1} \nonumber \\
& = \mathbf{G} \big(\mathbf{I}_{K} \otimes \mathbf{Q}_{P} \mathbf{C}_f \big) \bm{\Xi} \mathbf{x},
\end{align}
where $M \triangleq ON + \tfrac{N}{2}(K-1)$. 

\subsubsection{Additional Constraints}

The transmitted symbols are delayed from one another every $N/2$ samples, which is transparently ensured by the structure of $\mathbf{G}$ through the inclusion of $\mathbf{0}_{N/2}$ matrices. 
This representation as a sum of delayed matrices is detailed in \cite{pereira2022generalized}. 

It follows from the above that the proposed structure constitutes an affine-precoded filter-bank scheme, which can be implemented efficiently using a \ac{PPN} together with an \ac{IFFT}. 
Alternatively, the proposed waveform can be interpreted as a modification of \ac{AFDM}, where the standard sinc-chirp subcarriers are replaced with chirp-filtered subcarriers. 

In addition, the length $P$ of the \ac{IDAFT} at the transmitter must satisfy $L < P < N$. 
The condition $P > L$ ensures that the precoding stage (via \ac{DAFT}) is not nullified by the \ac{IDAFT} of the filter-bank structure, whereas $P < N$ guarantees that the chirps are sampled at a rate lower than the Nyquist rate, thereby enabling frequency containment.

\vspace{-1ex}
\subsection{Receive Signal Model}
\label{subsec:received_signal}

The transmit signal vector $\mathbf{s}$ in \eqref{eq:td_tx_signal} is propagated through a time-varying multipath channel, i.e., a doubly-dispersive channel, represented concisely by the circular convolutional matrix $\mathbf{H} \in \mathbb{C}^{(K-1)M \times (K-1)M}$ \cite{nissel2018pruned}. 
Consequently, the received signal $\mathbf{r} \in \mathbb{C}^{M \times 1}$ is expressed as
\begin{equation}
\mathbf{r} \triangleq \hspace{-3.5ex}
\underbrace{(\mathbf{I}_{K-1} \otimes \check{\mathbf{H}})}_{\triangleq \mathbf{H} \in \mathbb{C}^{(K-1)M \times (K-1)M}} \hspace{-3.5ex}
\mathbf{G} \Big(\mathbf{I}_{K} \otimes \mathbf{Q}_{P} \mathbf{C}_f\Big) \bm{\Xi} \mathbf{x} + \mathbf{n},
\end{equation}
where $\check{\mathbf{H}} \in \mathbb{C}^{M \times M}$ denotes the doubly-dispersive channel composed of $R$ resolvable paths. 
Each $r$-th path induces a delay $\tau_r \in [0, \tau^\mathrm{max}]$ and a Doppler shift $\nu_r \in [-\nu^\mathrm{max}, +\nu^\mathrm{max}]$, with normalized integer delay $\ell_r \triangleq \lfloor \tfrac{\tau_r}{T_\mathrm{s}} \rceil \in \mathbb{N}_0$ and normalized Doppler $f_r \triangleq \tfrac{N\nu_r}{f_\mathrm{s}} \in \mathbb{R}$, where $f_\mathrm{s} \triangleq \tfrac{1}{T_\mathrm{s}}$ denotes the sampling frequency. 
The noise vector $\mathbf{n} \in \mathbb{C}^{M \times 1}$ represents \ac{AWGN} samples with variance $\sigma_n^2$. 

For simplicity, it is assumed that the doubly-dispersive channel remains constant during the $K$ time slots.\footnote{This assumption allows the definition $\bar{N} = NK$ for the linear detection procedure in Section~\ref{secReceivercomm}. A time-varying case can also be considered but is left for future work.} 

As described in \cite{Rou_SPM_2024}, the channel matrix $\check{\mathbf{H}}$ is expressed as  \vspace{-1ex}
\begin{equation}
\check{\mathbf{H}} \triangleq \sum_{r=1}^{R} h_r \mathbf{\Phi}_{r} \mathbf{Z}^{f_r} \mathbf{\Pi}^{\ell_r} \in \mathbb{C}^{M \times M},
\vspace{-1ex}
\end{equation}
where $h_r \in \mathbb{C}$ denotes the complex fading coefficient of the $r$-th path, $\mathbf{\Phi}_r \in \mathbb{C}^{M \times M}$ is the IDAFT-based chirp-cyclic prefix phase matrix, $\mathbf{Z} \in \mathbb{C}^{M \times M}$ is the diagonal roots-of-unity matrix, and $\mathbf{\Pi} \in \mathbb{C}^{M \times M}$ is the circular left-shift matrix. 
The phase matrix is given by
\begin{equation}
\mathbf{\Phi}_{r} \triangleq \mathrm{diag}\big[e^{-j2\pi \cdot \phi(\ell_r)}, \ldots, e^{-j2\pi \cdot \phi(1)}, 1, \ldots, 1\big],
\label{eq:CCP_phase_matrix}
\end{equation}
with $\phi(m) \triangleq c_1(M^2 - 2Mm)$. 
The matrix $\mathbf{Z}$ is defined as
\begin{equation}
\mathbf{Z} \triangleq \mathrm{diag}\big[e^{-j2\pi \tfrac{0}{M}}, \ldots, e^{-j2\pi \tfrac{M-1}{M}}\big],
\label{eq:Z_matrix}
\end{equation}
and is raised to the power $f_r$. 
In total, $\check{\mathbf{H}}$ is formed by $R$ diagonals, with positions determined by the path delays and coefficients modulated by the Doppler shifts.

The received signal is next demodulated by $(\mathbf{G}\mathbf{Q})\herm$, producing $\mathbf{\tilde{a}} \in \mathbb{C}^{LK \times 1}$. 
Considering all time slots, the detected symbols $\mathbf{\tilde{B}} \in \mathbb{C}^{L \times K}$ are obtained through the \ac{IDAFT} combined with the compensation stage as
\begin{equation}
\mathbf{\tilde{B}} = \mathbf{W}\herm_L \mathrm{diag}\{\tilde{\mathbf{b}}\} \mathbf{\tilde{A}},
\label{12ghtj}
\end{equation}
where $\mathbf{\tilde{A}} = [\mathbf{\tilde{a}}_0;\mathbf{\tilde{a}}_1;\ldots;\mathbf{\tilde{a}}_{K-1}] \in \mathbb{C}^{L \times K}$. 
Since no data are transmitted in the intermediate $L/2$ positions of $\mathbf{A}$, these are discarded at the receiver. 
For symmetry, the compensation stage is applied at both transmitter and receiver, but it is effective only on one side. 

Concatenating all effects, the final received signal $\mathbf{y} \in \mathbb{C}^{K\frac{L}{2} \times 1}$ in the absence of noise is expressed as
\begin{equation}
\label{eq:final_IO}
\mathbf{y} \triangleq \bm{\Xi}\herm \Big(\mathbf{I}_{K} \otimes \mathbf{C}_f\herm \mathbf{Q}_{P}\herm \Big) \mathbf{G}\herm \mathbf{H} \mathbf{G} \Big(\mathbf{I}_{K} \otimes \mathbf{Q}_{P} \mathbf{C}_f\Big) \bm{\Xi} \mathbf{x},
\end{equation}
where the end-to-end effective channel matrix is defined as
\begin{equation}
\mathbf{H}_\text{eff} \triangleq \bm{\Xi}\herm \Big(\mathbf{I}_{K} \otimes \mathbf{C}_f\herm \mathbf{Q}_{P}\herm \Big) 
\underbrace{\mathbf{G}\herm \mathbf{H} \mathbf{G} \Big(\mathbf{I}_{K} \otimes \mathbf{Q}_{P} \mathbf{C}_f\Big) \bm{\Xi}}_{\bar{\mathbf{H}} \in \mathbb{C}^{NK \times K\frac{L}{2}}}.
\label{eq:Heff}
\end{equation}

\subsection{Waveform Complexity}

The overall computational complexity of the proposed waveform can be expressed as
\begin{eqnarray*}
&\hspace{-3ex}\mathcal{O}(ON \!+\! N \log N \!+\! N \log N \!+\! P \log P \!+\! 4L \!+\! 2L \log L \!+\! L) \\ &\approx
\mathcal{O}(ON + 5L + 2N \log N + 3L \log L),  
\end{eqnarray*} 
at the transmitter, where
\begin{itemize}
\item $ON + N \log N$ corresponds to the filter-bank modulator, implemented as an $N$-point IFFT combined with $ON$ prototype filter multiplications;
\item $N \log N + P \log P$ corresponds to the frequency-domain zero padding;
\item $4L + 2L \log L$ corresponds to the (I)DAFT operations, including spreading;
\item $L$ corresponds to the filter compensation stage.
\end{itemize}

The second line assumes $P \approx L$. 
The $K$-point (I)DFTs are assumed to require $K \log K$ multiplications. 
For reference, the overall complexity of the comparable DAFT-s-AFDM scheme with bandwidth control, as described in \cite{savaux2024}, is given by $2P + 3P \log P + N \log N$. 
Considering that the proposed waveform incorporates per-subcarrier filtering while achieving very low \ac{OOBE}, the additional computational complexity relative to \ac{AFDM} and its variants is regarded as manageable.

\section{Analysis of AFBM Channel Characteristics}
\label{sec:AFBM_channel_characteristics}

\subsection{Intermediate Effective Channel Analysis}

The demodulation and equalization of the filtered doubly-dispersive channel matrix for the \ac{AFBM} waveform can be expressed as \vspace{-1ex}
\begin{equation}
\acute{\mathbf{H}} \triangleq \mathbf{Q}_P^{H} \mathbf{G}^{H} \bigg(\sum_{r=1}^{R} h_r \mathbf{\Phi}_r \mathbf{Z}^{f_r} \mathbf{\Pi}^{\ell_r}\bigg) \mathbf{G} \mathbf{Q}_{P} \in \mathbb{C}^{L \times L},
\label{eq:paths_channel} \vspace{-1ex}
\end{equation}
which represents the structure of the doubly-dispersive channel after filtering and truncated IDAFT/DAFT processing, but before de-spreading and filter compensation. 
These latter operations are not considered here, since they do not affect the fundamental properties of the waveform over the core doubly-dispersive channel operations governed by the modulating DAFT/IDAFT matrix $\mathbf{Q}_P$.

\begin{figure}[H]
\centering		
\includegraphics[width=1\columnwidth]{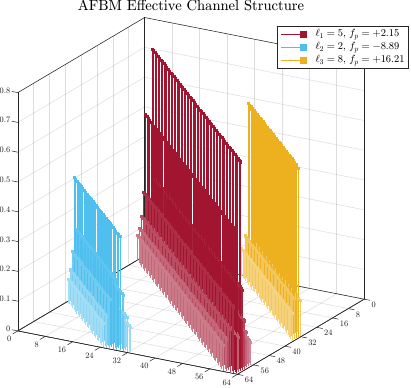}
\caption{3D illustration of a three-path \ac{AFBM} intermediate effective channel of the DAFT-spread signal. 
The normalized delay and Doppler shifts are indicated in the legend. 
The parameters are $L = 64$, $N = 128$, $P = 128$, $O = 4$, with the PHYDYAS4 prototype filter.}
\label{fig:AFBM_chan}
\end{figure}

This formulation is particularly useful for illustrating how the \ac{AFBM} waveform spreads the band-diagonals of the channel, a mechanism that plays a central role in improving robustness against inter-path interference. 
An example is shown in Fig.~\ref{fig:AFBM_chan}, where the intermediate effective channel of \ac{AFBM} is depicted for a scenario with three distinct propagation paths, and the spreading of the band-diagonals can be clearly observed.

This behavior closely resembles that of the \ac{AFDM} waveform \cite{bemani2023,Rou_SPM_2024}, as both are inherently shaped by the IDAFT-based modulation process. 
The main distinction arises from the filter design, which introduces a different sideband structure, but the fundamental outcome remains the deterministic shifting of the diagonals according to the path delay and Doppler indices. 
Such spreading is the key factor enabling robustness to inter-path interference and, at the same time, provides highly beneficial characteristics for \ac{ISAC} parameter estimation. 
In summary, the proposed \ac{AFBM} scheme inherits the robustness properties of \ac{AFDM} over doubly-dispersive channels while further enhancing its potential for joint communication and sensing.

\begin{figure*}[t!]
\centering		
\includegraphics[width=\textwidth]{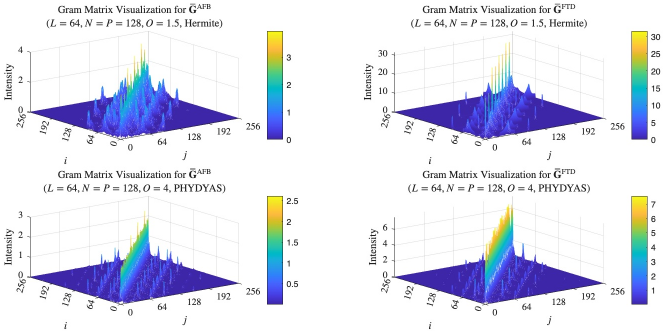}
\vspace{-3ex}
\caption{Illustration of the Gram matrix structure of the two effective channels with various prototype filter types computed via equations \eqref{eq:Gram_AFB} and \eqref{eq:Gram_FTD}.} 
\label{fig:AFBM_Gram}
\vspace{1ex}
\end{figure*}

\subsection{Gram Matrix Analysis for ISAC Receiver Design}
\label{subsec:Gram_matrix_analysis}

In this subsection, the rationale for adopting the hybrid filtered \ac{TD} effective \ac{I/O} relationship with $\bar{\mathbf{H}}$ in \eqref{eq:Heff} is explained, as opposed to using the full effective channel $\mathbf{H}_\text{eff}$ in \eqref{eq:Heff}, when Bayesian interference cancellation algorithms are executed for \ac{ISAC} receiver design. 

The proposed \ac{GaBP}-based communication receiver improves estimation accuracy iteratively by solving a linear regression problem based on a low-complexity \ac{MF}. 
Unlike the \ac{MMSE} filter, which explicitly requires a matrix inversion, the \ac{MF} alone cannot sufficiently suppress the correlation between observations. 
Therefore, in order to achieve accurate estimation, the filtering must be carried out in a signal domain where instantaneous correlations among observations are minimized \cite{TakahashiTWC2022}. 
The degree of correlation in different signal domains can be visually confirmed through the Gram matrix of the corresponding equivalent observation matrix \cite{TakahashiTCOM2019}.

Figure~\ref{fig:AFBM_Gram} compares the intensity distributions of the Gram matrices in two cases: the AFB-domain and the hybrid filtered \ac{TD} domain. 
These are defined, respectively, as
\begin{subequations}
\begin{equation}
\label{eq:Gram_AFB}
\bar{\mathbf{G}}^\text{AFB} \triangleq \mathbf{H}_\text{eff}\herm \mathbf{H}_\text{eff} \in \mathbb{C}^{K\frac{L}{2} \times K\frac{L}{2}},
\end{equation}
\begin{equation}
\label{eq:Gram_FTD}
\bar{\mathbf{G}}^\text{FTD} \triangleq \bar{\mathbf{H}}\herm \bar{\mathbf{H}} \in \mathbb{C}^{K\frac{L}{2} \times K\frac{L}{2}}.
\end{equation}
\end{subequations}

As observed in the upper part of Fig.~\ref{fig:AFBM_Gram}, the fixed correlation structure induced by $\bm{\Xi}\herm \big(\mathbf{I}_{K} \otimes \mathbf{C}_f\herm \mathbf{Q}_{P}\herm \big)$ emphasizes the off-diagonal entries of the AFB-domain Gram matrix in a regular pattern, leading to large instantaneous correlations among the receive observations. 
In contrast, the lower part of Fig.~\ref{fig:AFBM_Gram} shows that the hybrid filtered \ac{TD} Gram matrix approaches a diagonal form in high-dimensional \ac{AFBM} systems, a result of the increased randomness of the observation matrix. 
This property implies that, when designing low-complexity Bayesian algorithms based on the \ac{MF}, the hybrid filtered \ac{TD} Gram matrix $\bar{\mathbf{G}}^\text{FTD}$ should be employed rather than the AFB-domain Gram matrix $\bar{\mathbf{G}}^\text{AFB}$. 
This insight constitutes a key factor underlying the robustness of the proposed \ac{GaBP}- and \ac{PDA}-based receivers. 

\section{PAPR, OOBE, and AF Analysis}
\label{sec:PAPR_OOBE}

In this section, further characteristics of the proposed waveform are analyzed, focusing on \acf{PAPR}, \acf{OOBE}, and the shape of the \acf{AF}.

\subsection{PAPR}

\begin{figure}[H]
\vspace{-1ex}
\centering
\includegraphics[width=\columnwidth]{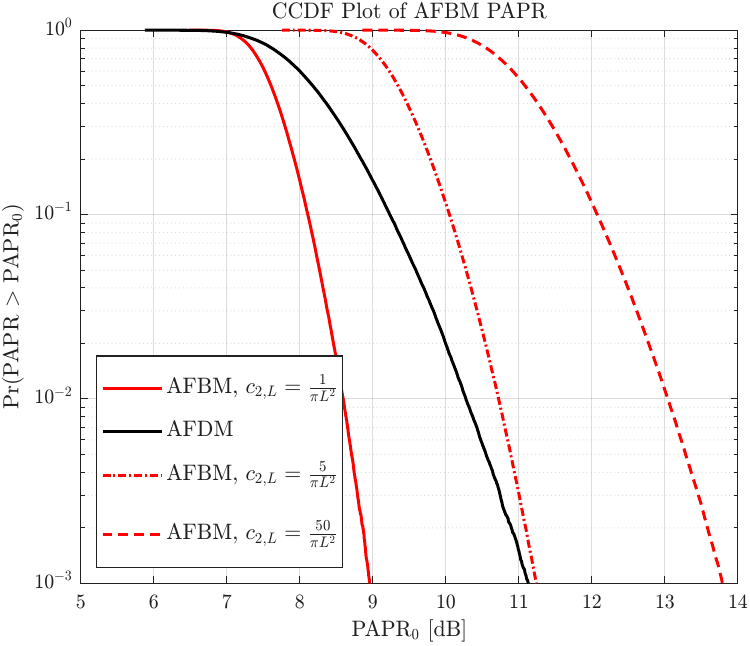}%
\caption{\ac{PAPR} performance of \ac{AFDM} and \ac{AFBM}.}
\label{fig:afdm_papr}
\end{figure}

Figure~\ref{fig:afdm_papr} illustrates the \ac{PAPR} performance of the considered systems. 
It can be observed that, due to its single-carrier-like structure, the proposed \ac{AFBM} waveform achieves an advantage of approximately 2~dB compared with regular \ac{AFDM}, which exhibits the same high \ac{PAPR} levels as conventional \ac{OFDM}-based schemes. 
Furthermore, since the proposed waveform can be regarded as an affine extension of the pruned \ac{DFT}-spread \ac{FBMC}, it preserves the low \ac{PAPR} properties of that scheme. 
As also reported for DAFT-s-AFDM in \cite{tao2024daft}, the selection of the chirp parameter $c_2$ is critical for maintaining low \ac{PAPR}. 
Specifically, the lowest values were obtained with $c_{2,L} = \tfrac{1}{\pi L^2}$, $c_{2,P} = \tfrac{1}{\pi P^2}$, and $c_{2,N} = \tfrac{1}{\pi N^2}$. 
When $c_{2,L}$ is increased to $\tfrac{5}{\pi L^2}$, the \ac{PAPR} approaches that of AFDM, while further increasing to $\tfrac{50}{\pi L^2}$ results in a \ac{PAPR} even higher than AFDM.

\begin{figure}[H]
\centering
\includegraphics[width=0.95\columnwidth]{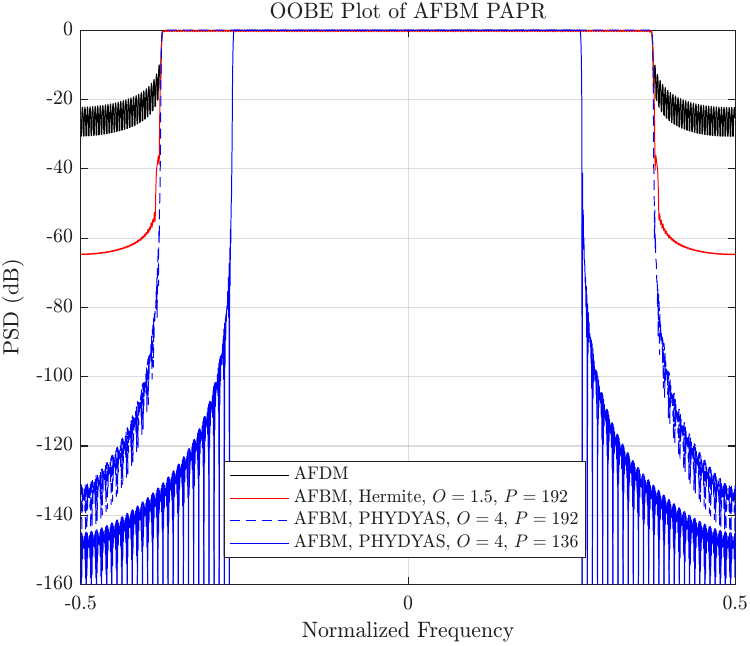}%
\vspace{-1.5ex}
\caption{\ac{OOBE} performance of \ac{AFDM} and \ac{AFBM} with the Hermite and PHYDYAS prototype filters.}
\label{fig:afdm_oob}
\vspace{-1.5ex}
\end{figure}

\subsection{OOBE}

The power spectral density $\mathbf{e} \in \mathbb{C}^{ON + (K-1)N/2}$ of the transmitted signal can be computed as \vspace{-0.5ex}
\begin{equation}
\tilde{\mathbf{e}} = \mathrm{diag}\left\{ \mathbf{Q}^{H}\mathbf{G}^{T}\mathbf{GQ} \right\}. \vspace{-0.5ex}
\end{equation}

The normalized \ac{OOBE} performance of the considered schemes is presented in Fig.~\ref{fig:afdm_oob}. 
Two prototype filters were employed for \ac{AFBM}: a truncated Hermite filter with overlap $O = 1.5$, and the PHYDYAS filter with overlap $O = 4$. 
By employing well-localized filters in place of a rectangular window, the proposed waveform achieves a significant improvement in \ac{OOBE} compared with regular \ac{AFDM}. 
An interesting tradeoff arises with respect to the choice of prototype filter: a shorter filter leads to reduced latency but weaker spectral localization, although still considerably better than that of AFDM, while a longer filter provides stronger suppression of out-of-band emissions. 
In addition, the parameter $P$ directly determines the occupied bandwidth and therefore influences the spectral efficiency of \ac{AFBM}, with the practical condition $P > L$ being recalled. 
Finally, when nonlinearities of power amplifiers are taken into account, the favorable \ac{OOBE} characteristics of the proposed waveform are not expected to degrade severely, since its inherently low \ac{PAPR} mitigates such effects. 

\vspace{-1ex}
\subsection{Ambiguity Function}

\begin{figure}[H]
\centering
\includegraphics[width=1\columnwidth]{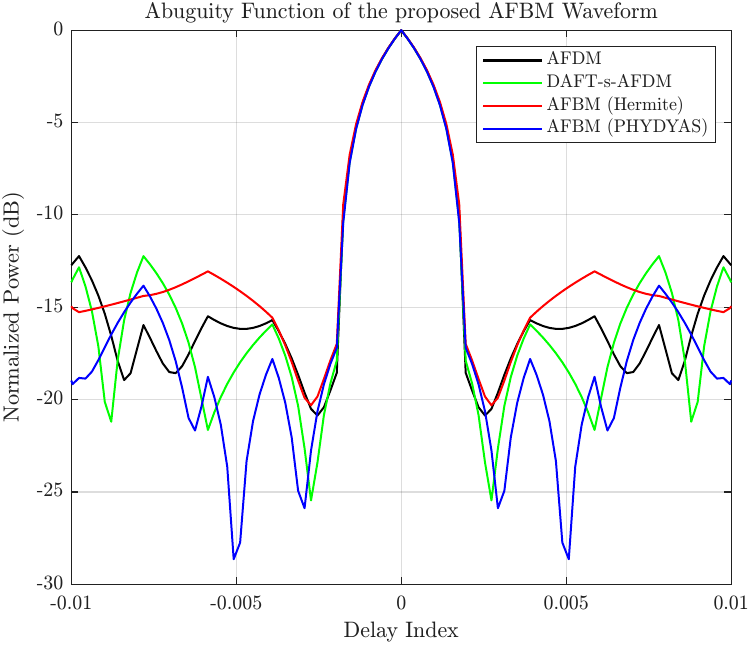}%
\vspace{-2ex}
\caption{Delay AF performance of \ac{AFDM} and \ac{AFBM} with the Hermite and PHYDYAS prototype filters.}
\label{fig:afdm_AF}
\vspace{-3ex}
\end{figure}

Finally, the ambiguity functions of the considered scheme are analyzed, as illustrated in Fig.~\ref{fig:afdm_AF}. 
In this scenario, two prototype filters are considered for \ac{AFBM}: a truncated Hermite filter with overlap $O = 1.5$ and the PHYDYAS filter with overlap $O = 4$. 
The ambiguity function of \ac{AFBM} is found to be very similar to that of \ac{AFDM}, with a slight improvement in sidelobe suppression observed when the PHYDYAS prototype filter is employed. 
Since the shape of the ambiguity function varies with the choice of the chirp parameter $c_2$, heuristic values were selected for each waveform to ensure the best performance.\footnote{Although optimization problems can, in principle, be formulated to further enhance certain desirable features, this line of work is left for future investigation.} 

\begin{figure}[H]
\centering
\vspace{1ex}
\includegraphics[width=1\columnwidth]{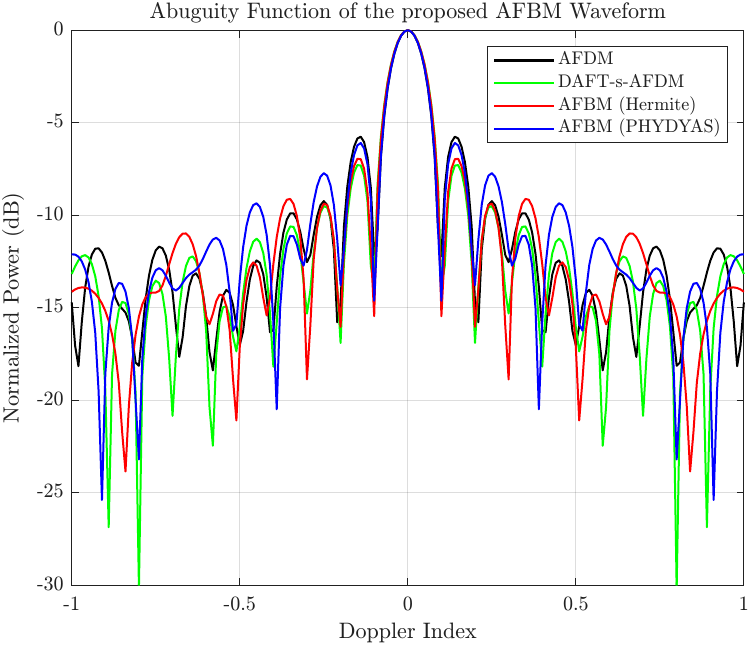}%
\vspace{-2ex}
\caption{Doppler AF performance of \ac{AFDM} and \ac{AFBM} with the Hermite and PHYDYAS prototype filters.}
\label{fig:afdm_AF_dop}
\end{figure}

The chosen parameters are as follows: $c_{2,\bar{M}} = \pi / \bar{M}^2$ for \ac{AFDM}; $c_{2,\bar{M}} = 3\text{e}100$ and $c_{2,\bar{D}} = 0$ for DAFT-s-AFDM\footnote{Here, the spreading parameter for DAFT-s-AFDM is defined as $\bar{D} \triangleq K\frac{L}{2}$.} and $c_{2,N} = 0$, $c_{2,L} = \pi / L^2$, and $c_{2,P} = 0$ for \ac{AFBM}.

\vspace{1ex}
\section{Proposed GaBP-based Receiver Design}
\label{secReceivercomm}

From a receiver design perspective, the goal is to estimate the transmitted signal $\mathbf{x}$ under the assumption that the filtered \ac{TD} channel matrix $\bar{\mathbf{H}} \in \mathbb{C}^{NK \times K\frac{L}{2}}$ is known.\footnote{The final \ac{I/O} relationship in \eqref{eq:final_IO} is not employed since the Gram matrix of $\mathbf{H}_\text{eff}$ does not approach a clear diagonal, as discussed in Section~\ref{subsec:Gram_matrix_analysis}.} 
Accordingly, the \ac{GaBP}-based detector for \ac{AFBM} is derived from the \ac{I/O} relationship
\begin{equation}
\label{General_I/O_arbitrary}
\bar{\mathbf{r}} = \bar{\mathbf{H}}  \mathbf{x} + \bar{\mathbf{w}}.
\end{equation}

For benchmarking, the corresponding high-complexity \ac{LMMSE} estimators in both the hybrid filtered \ac{TD} and AFB domains can be expressed, respectively, as
\vspace{-1ex}
\begin{subequations}
\begin{equation}
\label{eq:LMMSE}
\hat{\mathbf{x}}_\text{LMMSE}^\text{FTD} = \left( \bar{\mathbf{H}}\herm \bar{\mathbf{H}} + \sigma^2_n \mathbf{I}_{K\frac{L}{2}} \right)^{-1} \bar{\mathbf{H}}\herm \bar{\mathbf{r}},
\end{equation}
\begin{equation}
\label{eq:LMMSE_AFB}
\hat{\mathbf{x}}_\text{LMMSE}^\text{AFB} = \left( \mathbf{H}_\text{eff}\herm \mathbf{H}_\text{eff} + \sigma^2_n \mathbf{I}_{K\frac{L}{2}} \right)^{-1} \mathbf{H}_\text{eff}\herm \mathbf{y}.
\end{equation}
\end{subequations}

It should be noted that, with this formulation, no additional demodulation stage is required once the effects of $\mathbf{G}$ have been reversed for data detection.

\subsection{GaBP-based Data Detection}

Let $\bar{N} \triangleq NK$ and $\bar{M} \triangleq K\frac{L}{2}$, with index sets $\bar{n} \triangleq \{1,\dots,\bar{N}\}$ and $\bar{m} \triangleq \{1,\dots,\bar{M}\}$. 
The element-wise representation of \eqref{General_I/O_arbitrary} is given by
\vspace{-1ex}
\begin{equation}
\label{General_I/O_arbitrary_elementwise}
\bar{r}_{\bar{n}} = \sum_{\bar{m}=1}^{\bar{M}} \bar{h}_{\bar{n},\bar{m}} x_{\bar{m}} + \bar{w}_{\bar{n}}, 
\vspace{-1ex}
\end{equation}
so that the soft replica of the $\bar{m}$-th communication symbol associated with the $\bar{n}$-th received signal $\bar{r}_{\bar{n}}$ at the $i$-th iteration of a message-passing algorithm is denoted by $\hat{x}_{\bar{n},\bar{m}}^{(i)}$. 
The corresponding \ac{MSE} is computed as
\vspace{-0.5ex}
\begin{equation}
\hat{\sigma}^{2(i)}_{x:{\bar{n},\bar{m}}} \triangleq \mathbb{E}_{x} \big[ | x - \hat{x}_{\bar{n},\bar{m}}^{(i-1)} |^2 \big]= E_\mathrm{S} - |\hat{x}_{\bar{n},\bar{m}}^{(i-1)}|^2, \forall (\bar{n},\bar{m}),
\label{eq:MSE_d_k}
\vspace{-0.5ex}
\end{equation}
where $\mathbb{E}_{x}$ denotes the expectation over all possible transmitted symbols $x$.

The proposed \ac{GaBP} receiver proceeds in three major stages, described below.

\subsubsection{Soft Interference Cancellation}

At the $i$-th iteration, the soft replicas $\hat{x}_{\bar{n},\bar{m}}^{(i-1)}$ obtained in the previous iteration are used to compute the data-centric \ac{sIC} signals $\tilde{r}_{x:\bar{n},\bar{m}}^{(i)}$. 
From \eqref{General_I/O_arbitrary_elementwise}, these are given as
\vspace{-1ex}
\begin{align}
\label{eq:d_soft_IC}
\tilde{r}_{x:\bar{n},\bar{m}}^{(i)} &= \bar{r}_{\bar{n}} - \sum_{e \neq \bar{m}} h_{\bar{n},e} \hat{x}_{\bar{n},e}^{(i)}, \\
&= h_{\bar{n},\bar{m}} x_{\bar{m}} + \underbrace{\sum_{e \neq \bar{m}} h_{\bar{n},e}(x_e - \hat{x}_{\bar{n},e}^{(i)}) + \bar{w}_{\bar{n}}}_\text{interference + noise term}.
\end{align}

By invoking the \ac{SGA}, the interference and noise term is approximated as Gaussian, which yields the conditional \acp{PDF}
\vspace{-0.5ex}
\begin{equation}
\label{eq:cond_PDF_d}
p_{\tilde{r}_{x:\bar{n},\bar{m}}^{(i)} \mid x_{\bar{m}}}(\tilde{r}_{x:\bar{n},\bar{m}}^{(i)}|x_{\bar{m}}) \propto \exp\!\Bigg[ -\frac{|\tilde{r}_{x:\bar{n},\bar{m}}^{(i)} - h_{\bar{n},\bar{m}} x_{\bar{m}}|^2}{\tilde{\sigma}_{x:\bar{n},\bar{m}}^{2(i)}} \Bigg],
\vspace{-0.5ex}
\end{equation}
with conditional variances
\vspace{-0.5ex}
\begin{equation}
\label{eq:soft_IC_var_d}
\tilde{\sigma}_{x:\bar{n},\bar{m}}^{2(i)} = \sum_{e \neq \bar{m}} \left|h_{\bar{n},e}\right|^2 \hat{\sigma}^{2(i)}_{x:{\bar{n},e}} + \sigma^2_n.
\end{equation}

\subsubsection{Belief Generation}

In the belief generation stage, the \ac{SGA} is again invoked, under the assumptions that $\bar{N}$ is sufficiently large and that the estimation errors in $\hat{x}_{\bar{n},\bar{m}}^{(i-1)}$ are independent, in order to construct initial symbol estimates (beliefs). 
Using the conditional \acp{PDF} in \eqref{eq:cond_PDF_d}, the following extrinsic distributions are obtained:
\begin{equation}
\label{eq:extrinsic_PDF_d}
\prod_{e \neq \bar{n}} p_{\tilde{r}_{x:e,\bar{m}}^{(i)} \mid x_{\bar{m}}}(\tilde{r}_{x:e,\bar{m}}^{(i)}|x_{\bar{m}}) \propto \exp\!\Bigg[ - \frac{(x_{\bar{m}} - \bar{x}_{\bar{n},\bar{m}}^{(i)})^2}{\bar{\sigma}_{x:\bar{n},\bar{m}}^{2(i)}} \Bigg],
\end{equation}
where the extrinsic means and variances are given by
\vspace{-1ex}
\begin{equation}
\label{eq:extrinsic_mean_d}
\bar{x}_{\bar{n},\bar{m}}^{(i)} = \bar{\sigma}_{x:\bar{n},\bar{m}}^{(i)} \sum_{e \neq \bar{n}} \frac{h^*_{e,\bar{m}} \tilde{r}_{x:e,\bar{m}}^{(i)}}{ \tilde{\sigma}_{x:e,\bar{m}}^{2(i)}},
\end{equation}
\vspace{-1ex}
\begin{equation}
\label{eq:extrinsic_var_d}
\bar{\sigma}_{x:\bar{n},\bar{m}}^{2(i)} = \bigg( \sum_{e \neq \bar{n}} \frac{|h_{e,\bar{m}}|^2}{\tilde{\sigma}_{x:e,\bar{m}}^{2(i)}} \bigg)^{-1},
\end{equation}
with $h^*_{e,\bar{m}}$ denoting the complex conjugate of $h_{e,\bar{m}}$.

\subsubsection{Soft Replica Generation}

In the final stage, the beliefs are denoised according to a Bayes-optimal rule to obtain the refined symbol estimates. 
For \ac{QPSK} modulation,\footnote{\ac{QPSK} is considered here without a loss of generality, as they nicely admit to a closed-form denoiser, but denoisers for higher-order constellations can of course be designed \cite{TakahashiTCOM2019}.} the Bayes-optimal denoiser is expressed as
\vspace{-1ex}
\begin{equation}
\hat{x}_{\bar{n},\bar{m}}^{(i)} = c_x \Bigg( \tanh\!\Big[ 2c_d \frac{\Re\{\bar{x}_{\bar{n},\bar{m}}^{(i)}\}}{\bar{\sigma}_{x:{\bar{n},\bar{m}}}^{2(i)}} \Big] + j \tanh\!\Big[ 2c_d \frac{\Im\{\bar{x}_{\bar{n},\bar{m}}^{(i)}\}}{\bar{\sigma}_{x:{\bar{n},\bar{m}}}^{2(i)}} \Big] \Bigg),
\label{eq:QPSK_denoiser}
\end{equation}
where $c_x \triangleq \sqrt{E_\mathrm{S}/2}$ denotes the magnitude of the real and imaginary parts of the \ac{QPSK} symbols, with variances updated using \eqref{eq:MSE_d_k}.

To avoid convergence to local minima caused by incorrect hard-decision replicas \cite{Su_TSP_2015}, the outputs $\hat{x}_{\bar{n},\bar{m}}^{(i)}$ are damped with a factor $0 < \beta_x < 1$:
\vspace{-1ex}
\begin{equation}
\label{eq:d_damped}
\hat{x}_{\bar{n},\bar{m}}^{(i)} = \beta_x \hat{x}_{\bar{n},\bar{m}}^{(i)} + (1 - \beta_x) \hat{x}_{\bar{n},\bar{m}}^{(i-1)}.
\end{equation}

The variances $\hat{\sigma}^{2(i)}_{x:{\bar{n},\bar{m}}}$ are updated similarly, i.e.,
\vspace{-1ex}
\begin{equation}
\label{eq:MSE_d_m_damped}
\hat{\sigma}^{2(i)}_{x:{\bar{n},\bar{m}}} = \beta_x \hat{\sigma}_{x:{\bar{n},\bar{m}}}^{2(i)} + (1-\beta_x) \hat{\sigma}_{x:{\bar{n},\bar{m}}}^{2(i-1)}.
\end{equation}

Finally, to resolve the conflicting dimensions, the consensus update for the symbol estimates is obtained as
\vspace{-1ex}
\begin{equation}
\label{eq:d_hat_final_est}
\hat{x}_{\bar{m}} = \bigg( \sum_{\bar{n}=1}^{\bar{N}} \frac{|h_{\bar{n},\bar{m}}|^2}{\tilde{\sigma}_{x:\bar{n},\bar{m}}^{2(i_\text{max})}} \bigg)^{-1} 
\bigg( \sum_{\bar{n}=1}^{\bar{N}} \frac{h^*_{\bar{n},\bar{m}} \tilde{r}_{x:\bar{n},\bar{m}}^{(i_\text{max})}}{ \tilde{\sigma}_{x:\bar{n},\bar{m}}^{2(i_\text{max})}} \bigg).
\end{equation}

The complete pseudocode of the proposed detection procedure is summarized in Algorithm~\ref{alg:proposed_decoder} on the following page.

\subsection{Complexity Analysis}
\vspace{-1ex}

The complexity of the proposed \ac{GaBP} detection algorithm is linear on the number of element-wise operations, and its per-iteration computational complexity is given by $\mathcal{O}(\bar{N}\bar{M})$. 
Notice that this complexity is much lesser than that of typical detection methods such as the \ac{LMMSE}, which is $\mathcal{O}(\bar{M}^3)$ due to the costly matrix inversion involved.

\begin{algorithm}[H]
\caption{GaBP-based Data Detection for AFBM Systems}
\label{alg:proposed_decoder}
\setlength{\baselineskip}{11pt}
\textbf{Input:} Receive signal vector $\bar{\mathbf{r}}\in\mathbb{C}^{\bar{N}\times 1}$, complex channel matrix $\bar{\mathbf{H}} \in \mathbb{C}^{\bar{N}\times \bar{M}}$, number of \ac{GaBP} iterations $i_{\max}$, constellation power $E_\mathrm{S}$, noise power $\sigma^2_n$ and damping factor $\beta_x$. \\
\textbf{Output:} Estimated symbol vector $\hat{\mathbf{x}}$ 
\vspace{-2ex} 
\begin{algorithmic}[1]  
\STATEx \hspace{-3.5ex}\hrulefill
\STATEx \hspace{-3.5ex}\textbf{Initialization}
\STATEx \hspace{-3.5ex} - Set iteration counter to $i=0$ and amplitudes $c_x = \sqrt{E_\mathrm{S}/2}$.
\STATEx \hspace{-3.5ex} - Set initial data estimates to $\hat{x}_{\bar{n},\bar{m}}^{(0)} = 0$ and corresponding 
\STATEx \hspace{-2ex} variances to $\hat{\sigma}^{2(0)}_{x:{\bar{n},\bar{m}}} = E_\mathrm{S}, \forall \bar{n},\bar{m}$. \vspace{-1ex}
\STATEx \hspace{-3.5ex}\hrulefill
\STATEx \hspace{-3.5ex}\textbf{for} $i=1$ to $i_\text{max}$ \textbf{do}: $\forall \bar{n}, \bar{m}$
\STATE Compute \ac{sIC} data signal $\tilde{r}_{x:{\bar{n},\bar{m}}}^{(i)}$ and its corresponding variance $\tilde{\sigma}^{2(i)}_{x:{\bar{n},\bar{m}}}$ from equations \eqref{eq:d_soft_IC} and \eqref{eq:soft_IC_var_d}.
\STATE Compute extrinsic data signal belief $\bar{x}_{\bar{n},\bar{m}}^{(i)}$ and its corresponding variance $\bar{\sigma}_{x:{\bar{n},\bar{m}}}^{2(i)}$ from equations \eqref{eq:extrinsic_mean_d} and \eqref{eq:extrinsic_var_d}.
\STATE Compute denoised and damped data signal estimate $\hat{x}_{\bar{n},\bar{m}}^{(i)}$ from equations \eqref{eq:QPSK_denoiser} and \eqref{eq:d_damped}.
\STATE Compute denoised and damped data signal variance $\hat{\sigma}_{x:{\bar{n},\bar{m}}}^{2(i)}$ from equations \eqref{eq:MSE_d_k} and \eqref{eq:MSE_d_m_damped}.

\STATEx \hspace{-3.5ex}\textbf{end for}
\STATE Calculate $\hat{x}_{\bar{m}}, \forall \bar{m}$ (equivalently $\hat{\mathbf{x}}$) using equation \eqref{eq:d_hat_final_est}. 

\end{algorithmic}
\end{algorithm}

\vspace{-4ex}
\subsection{Simulation Results}
\vspace{-1ex}

For the simulations, the number of subcarriers was set to $L = 128$, and the filter-bank \ac{DFT} size was chosen as $N = 256$. 
Each transmission consisted of $K = 8$ symbols, with a carrier frequency of $f_c = 4$~GHz and a bandwidth of 1~MHz. 
The channel was modeled as a doubly dispersive channel with three resolvable propagation paths characterized by normalized delays and digital Doppler shifts. 
The chirp frequencies for each \ac{DAFT} and \ac{IDAFT} were selected to satisfy the orthogonality condition \cite{Rou_SPM_2024}, expressed as
\[
2(f^{\text{max}} + \xi)(\ell^{\text{max}}+1) + \ell^{\text{max}} \leq P,
\]
where $f^{\text{max}}$ and $\ell^{\text{max}}$ denote the maximum normalized digital Doppler shift and delay of the channel, respectively, and $\xi \in \mathbb{N}_0$ is a free parameter that determines the guard width, i.e., the number of additional guard elements placed around the diagonals to account for Doppler-domain interference. 
The channel parameters were chosen as $\ell^{\text{max}} = 16$ with randomly chosen integer values and $f^{\text{max}} = 2$ with randomly chosen fractional values, while the algorithmic parameters were set to $i_{\max} = 20$ and $\beta_x = 0.5$.

\begin{figure}[H]
\centering
\includegraphics[width=0.975\columnwidth]{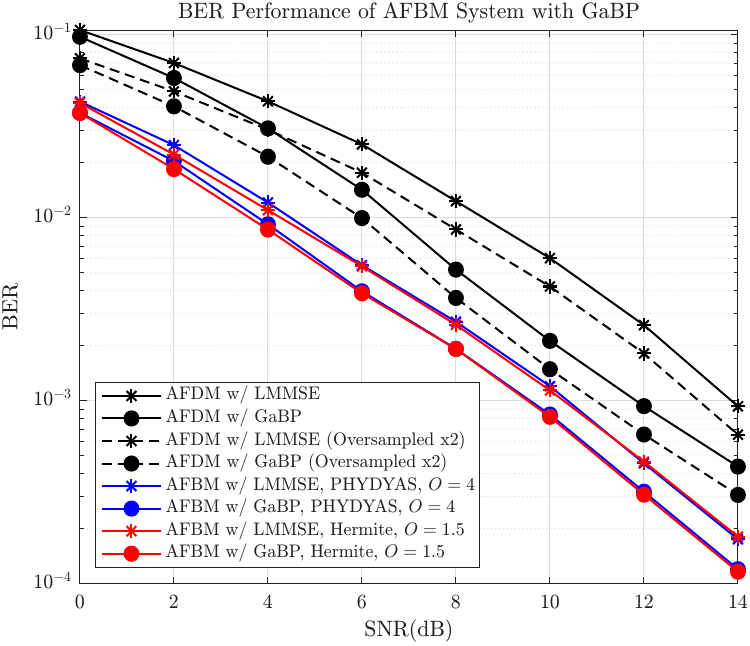}%
\vspace{-2ex}
\caption{\ac{BER} performance of the proposed \ac{GaBP} technique for both the \ac{AFDM} and \ac{AFBM} waveforms with Hermite and PHYDYAS prototype filters.}
\label{fig:afdm_ber}
\vspace{3ex}
\includegraphics[width=0.975\columnwidth]{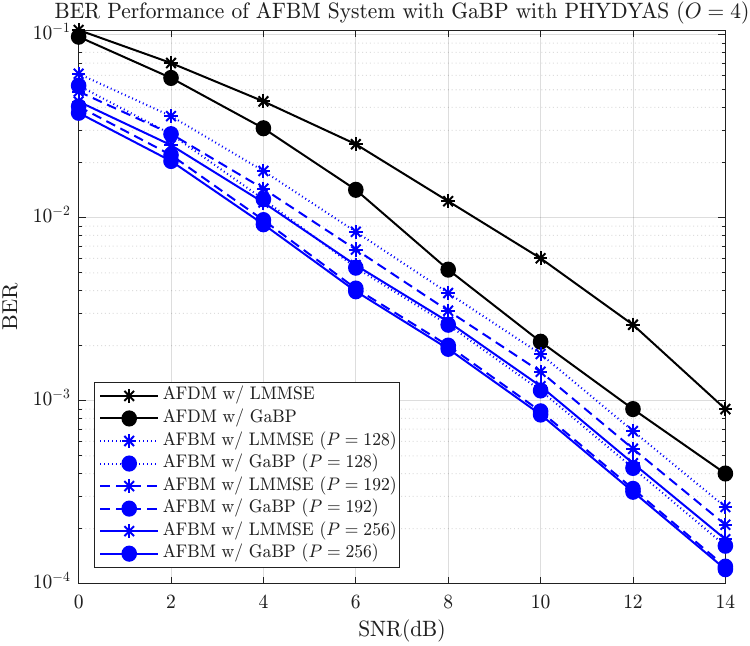}%
\vspace{-2ex}
\caption{\ac{BER} performance of the proposed \ac{GaBP} technique for both the \ac{AFDM} with PHYDYAS prototype filters for varying $P$.}
\label{fig:afdm_ber_p}
\end{figure}

Figures~\ref{fig:afdm_ber} and \ref{fig:afdm_ber_p} present the \ac{BER} performance of the considered systems. 
It is noted that, in contrast to the \ac{TD} relationship in \eqref{General_I/O_arbitrary} used for the proposed \ac{AFBM} scheme, the conventional \ac{DAF}-domain \ac{I/O} relationship reported in \cite{RanasingheTWC2025} was employed for \ac{AFDM}, since the effective Gram matrix structures are almost identical. 

In Fig.~\ref{fig:afdm_ber}, the chirp size was set to $P = 256$, representing the best-case scenario. 
The results show that the proposed \ac{GaBP}-based \ac{AFBM} scheme outperforms conventional \ac{AFDM} by approximately 2~dB at a \ac{BER} of $10^{-3}$, which can be attributed to the improved spectral localization and reduced inter-subcarrier interference achieved by the proposed waveform. 
%
%
Moreover, even when employing standard hybrid \ac{TD} \ac{LMMSE} detection, \ac{AFBM} achieves lower error rates compared with the corresponding \ac{AFDM} case.

The error performance is also nearly identical when either the Hermite or PHYDYAS prototype filters are used, which demonstrates that excellent \ac{OOBE} suppression can be achieved without sacrificing \ac{BER} performance. 

To examine scenarios with tighter frequency containment, Fig.~\ref{fig:afdm_ber_p} presents the \ac{BER} performance for different values of $P$. 
Although the error performance degrades as $P$ is reduced, the proposed \ac{AFBM} scheme consistently outperforms conventional \ac{AFDM}, even in the worst-case setting with $P = L$.

\section{Proposed PDA-based ISAC Receiver Design}
\label{secReceiverISAC}

In this section, we develop a \ac{PDA}-based \ac{RPE} algorithm enhanced with \ac{EM}.
Due to the properties of the Gram matrix described in Section \ref{sec:AFBM_channel_characteristics}, we adopt the unique filtered hybrid \ac{TD} model described in equation \eqref{General_I/O_arbitrary} to design the estimation scheme.

Let us first expand out the \ac{I/O} relationship in equation \eqref{General_I/O_arbitrary} to expose the complex channel coefficient as
\begin{equation}
\label{eq:sensing_IO_1}
\bar{\mathbf{r}} = \sum_{r=1}^{R}h_r (\underbrace{\mathbf{G}\herm \mathbf{\Phi}_r\mathbf{Z}^{f_r} \mathbf{\Pi}^{\ell_r} \mathbf{G} (\mathbf{I}_{K} \otimes \mathbf{Q}_{P} \mathbf{C}_f) \bm{\Xi}}_{\tilde{\mathbf{H}}_r \;\in\; \mathbb{C}^{NK \times K\frac{L}{2}}})  \mathbf{x} + \bar{\mathbf{w}}.
\end{equation}

Next, following \cite{RanasingheTWC2025}, we consider a typical \ac{DD} channel characterized by a maximum normalized delay spread $\text{max}(\ell_r)$ and a corresponding digital normalized Doppler spread $\text{max}(f_r)$, satisfying the relations $\text{max}(\ell_r) << M$ and $\text{max}(f_r) << M$.

Then, leveraging the properties defined above and defining $K_\tau \triangleq \{0,\cdots,k,\cdots,K_\tau-1\}$ and $D_\nu \triangleq \{0,\cdots,d,\cdots,D_\nu-1\}$ to be sets large enough to define a sufficiently fine grid discretizing the region determined by the maximum delay and Doppler values, the discretized channel matrix $\tilde{\mathbf{H}}_{k,d} \;\in\; \mathbb{C}^{NK \times K\frac{L}{2}}$ expressed in terms of its discrete delay and Doppler indices as opposed to the $R$ paths in $\tilde{\mathbf{H}}_r$ within equation \eqref{eq:sensing_IO_1} can be expressed as
\begin{equation}
\label{eq:DD_channel_definition_frac_Dopp}
\tilde{\mathbf{H}}_{k,d} = \mathbf{G}\herm \mathbf{\Phi}_k \mathbf{Z}^{f_d} \mathbf{\Pi}^{\ell_k} \mathbf{G} (\mathbf{I}_{K} \otimes \mathbf{Q}_{P} \mathbf{C}_f) \bm{\Xi}.
\end{equation}

This lets us express the \ac{I/O} relationship in equation \eqref{eq:sensing_IO_1} as
\begin{subequations}
\begin{equation}
  \label{eq:sensing_IO_2}
\bar{\mathbf{r}} \!=\!\! \sum_{k=0}^{K_\tau\! -\! 1} \sum_{d=0}^{D_\nu - 1}\!\! \underbrace{\tilde{\mathbf{H}}_{k,d} \cdot \mathbf{x}}_{\mathbf{e}_{k,d}\in \mathbb{C}^{\bar{N}\times 1}}\!\!\! \cdot h_{k,d} + \bar{\mathbf{w}} = \mathbf{E} \cdot \mathbf{h} + \bar{\mathbf{w}} \in \mathbb{C}^{\bar{N}\times 1},
\end{equation}
where we implicitly defined the vectors $\mathbf{e}_{k,d}\triangleq\mathbf{\Gamma}_{k,d} \cdot \mathbf{x}$, and explicitly identified the dictionary matrix $\mathbf{E} \in \mathbb{C}^{N\times K_\tau D_\nu}$, and sparse channel vector $\mathbf{h} \in \mathbb{C}^{K_\tau D_\nu \times 1}$, respectively defined as
%
\begin{equation}
\label{eq:sparse_dict_matrix_def}
\mathbf{E}\! \triangleq\! [\mathbf{e}_{0,0}, \dots, \mathbf{e}_{0,D_\nu\! -\! 1}, \dots, \mathbf{e}_{K_\tau\! -\! 1,0}, \dots, \mathbf{e}_{K_\tau\! -\! 1,D_\nu \!-\! 1}],
\vspace{-1ex}
\end{equation}
\begin{equation}
\label{eq:sparse_channel_vector}
\mathbf{h} \!\triangleq\! [h_{0,0}, \dots, h_{0,D_\nu\! -\! 1}, \dots, h_{K_\tau\! -\! 1,0}, \dots, h_{K_\tau\! -\! 1,D_\nu\! -\! 1}]\trans\!\!\!.
\end{equation}
\end{subequations}

Notice that if the resolution of such a grid is sufficiently fine, the only non-zero terms of the sum in equation \eqref{eq:DD_channel_definition_frac_Dopp} are those where both $\tau_k\approx \tau_p$ and $\nu_d \approx \nu_p$, which in turn implies that estimating these radar parameters amounts to estimating the $P$ channel gains such that $h_{k,d}\neq 0$.

\subsection{Proposed Method: PDA-based Message Passing Estimator}
\label{sec:Proposed_PDA_MP}

In this section, we introduce a new \ac{PDA}-based \ac{RPE} framework, developed under the premise that the prior distributions of the entries $\hat{h}_m$ within the sparse channel estimate vector $\mathbf{\hat{;h}}$ follow a Bernoulli-Gaussian model.
Equivalently, at the $i$-th iteration of the proposed algorithm, we represent the unknown channel to be estimated as
\begin{equation}
\label{eq:h_m_estimate_dist}
\hat{h}_m^{(i)} \sim p_{\,\text{h}_m} (\text{h}_m;\bm{\theta}^{(i)}),
\vspace{-2ex}
\end{equation}
with
\vspace{-0.5ex}
\begin{equation}
\label{eq:h_m_pdf}
\!\!p_{\,\text{h}_m} (\text{h}_m;\bm{\theta}^{(i)}) \triangleq\! (1\! -\! {\rho}^{(i)}) \delta(h_m)\! + \!{\rho}^{(i)} \mathcal{CN}\big(h_m;{\bar{h}}^{(i)},{\bar{\sigma}}^{(i)} \big),\!\!
\vspace{-0.5ex}
\end{equation}
where $\bm{\theta}^{(i)} \triangleq [{\rho}^{(i)}, {\bar{h}}^{(i)},{\bar{\sigma}}^{(i)}]$ denotes all three parameters of the distribution, namely, sparsity rate, mean and variance.

As indicated by the notation, the parameter vector $\bm{\theta}^{(i)}$ is updated in an iterative manner. In a similar fashion to the \ac{SBL} technique presented in \cite{Mehrotra_TCom_2023}, this update can be carried out using the \ac{EM} algorithm.
In contrast to the above-mentioned \ac{SotA} approach, the proposed method computes the estimates of $\mathbf{h}$ through a message-passing algorithm, which will be detailed in the following discussion.

For ease of reference, we introduce the notion of a soft replica (i.e., a tentative estimate) of $h_m$, denoted as  $\big\{ \hat{h}_{m}^{(i)} \big\} $, for which the corresponding \ac{MSE} is expressed as
\vspace{-0.5ex}
\begin{equation}
\hat{\sigma}_{h:m}^{2(i)} \triangleq \mathbb{E}_{\text{h}_m} \big[  |h_m - \hat{h}_{m}^{(i)}|^2   \big].
\label{eq:MSE_h_PDA}
\end{equation}

\subsubsection{Soft IC}

The \ac{sIC} expression corresponding to an estimate of $h_m$ can be expressed as
\vspace{-2ex}
\begin{equation}
\label{eq:Soft_IC_PDA}
\tilde{\mathbf{r}}_{h:m}^{(i)}\! = \bar{\mathbf{r}}\! -\!\!\! \sum_{q\neq m}\!\! \mathbf{e}_{q} \hat{h}_{q}^{(i)}  \!=\! \mathbf{e}_{m}h_m \!+\!\!\!\!\!\!\! \overbrace{\sum_{q \neq m}^{K_\tau D_\nu}  (\mathbf{e}_{q} h_q\! -\! \mathbf{e}_{q} \hat{h}_{q}^{(i)})\! +\! \tilde{\mathbf{w}}}^{\text{residual interference+noise component}}\!\!\!, \!\!
\end{equation}
where $\mathbf{e}_m$ is the $m$-th column of the dictionary matrix $\mathbf{E}$.

By invoking the \ac{CLT}, it follows that under large-system assumptions, the residual interference-plus-noise term can be well-approximated as a multivariate complex Gaussian random variable.
Equivalently, this justifies the application of the \ac{VGA}, whereby the conditional \ac{PDF} of the beliefs $\tilde{\mathbf{r}}_{h:m}^{(i)}$, conditioned on $h_m$, can be written as
\vspace{-0.5ex}
\begin{equation}
\label{eq:VGA_y_given_h}
\!\!\tilde{\mathbf{r}}_{h:m}^{(i)}\!\!\sim\! p_{\textbf{y} | \text{h}_{m}} \!(\textbf{y} | h_{m}\!) 
\!\propto\! \text{exp} \!\Big[\! -\! \big(\! \textbf{y}\! -\! \mathbf{e}_{m} h_{m}\! \big)\herm \mathbf{\Sigma}^{-1(i)}_{m}\! \big(\! \textbf{y} \!- \!\mathbf{e}_{m} h_{m} \!\big)\! \Big]\!,\!\!\!
\end{equation}
where $\textbf{y}$ is an auxiliary variable, and the conditional covariance matrix $\mathbf{\Sigma}_m^{(i)}$ is given by
\vspace{-1ex}
\begin{align}
\mathbf{\Sigma}_m^{(i)} & \triangleq \mathbb{E}_{\textbf{h,$\tilde{\mathbf{w}}$}|\hat{h}_m \neq h_m} \bigg[ \big( \tilde{\mathbf{r}}_{h:m}^{(i)} - \mathbf{e}_{m} h_{m} \big) \big( \tilde{\mathbf{r}}_{h:m}^{(i)} - \mathbf{e}_{m} h_{m} \big)\herm  \bigg]
\nonumber \\[-1ex]
& = \sum_{q \neq m}^{K_\tau D_\nu} \hat{\sigma}_{h:q}^{2(i)} \mathbf{e}_q \mathbf{e}_q\herm + N_0 \mathbf{I}_N,
\label{eq:covariance_matrix_PDA}
\end{align}
with $N_0$ denoting the noise power.

\subsubsection{Belief Generation}
The belief corresponding to the estimate of the $m$-th channel coefficient $h_m$ is derived by aggregating the contributions of all \ac{sIC} beliefs $\tilde{\mathbf{r}}_{h:m}^{(i)}$, according to the \ac{PDF}
\begin{equation}
p_{\text{h} | h_m} (\text{h} | h_m) \propto \text{exp} \Big[ - \tfrac{|\text{h} - \tilde{h}_m^{(i)}|^2}{\tilde{\sigma}_{\tilde{h}:m}^{2(i)}} \Big],
\label{eq:ell_extrinsic_belief_PDA}
\end{equation}
which yields
\begin{subequations}
\label{eq:mean_and_var_extrinsic_belief_PDA}
\begin{equation}
\tilde{h}_m^{(i)} \triangleq \frac{1}{\eta_m^{(i)}} \mathbf{e}_m\herm \mathbf{\Sigma}^{-1(i)} \tilde{\mathbf{r}}_{h:m}^{(i)}\;\;\text{and}\;\;
\tilde{\sigma}_{\tilde{h}:m}^{2(i)} \triangleq \frac{1 - \eta_m^{(i)} \hat{\sigma}_{h:m}^{2(i)}}{\eta_m^{(i)}},
\end{equation}
where $\eta_m^{(i)}$ is a normalization factor defined as
\begin{equation}
\label{eq:eta_PDA}
\eta_m^{(i)} \triangleq \mathbf{e}_m\herm \mathbf{\Sigma}^{-1(i)} \mathbf{e}_m,
\end{equation}
and the common conditional covariance matrix\footnote{The matrix inversion lemma \cite{Ito_ICC_2021} is used in the derivation of equation \eqref{eq:mean_and_var_extrinsic_belief_PDA} and by consequence \eqref{eq:eta_PDA}, such that the same inverse matrix $\mathbf{\Sigma}^{(i)}$ can be used instead of $\mathbf{\Sigma}_m^{(i)}$.}
is given by
\begin{equation}
\mathbf{\Sigma}^{(i)} \triangleq \sum_{m=1}^{K_\tau D_\nu} \hat{\sigma}_{h:m}^{2(i)} \mathbf{e}_m \mathbf{e}_m\herm + N_0 \mathbf{I}_N.
\end{equation}
\end{subequations}

\subsubsection{Soft RG}

The soft replicas of $h_m$ are obtained by evaluating the conditional expectation with respect to the extrinsic beliefs, under the assumption that the effective noise terms in $\hat{h}_m^{(i)}, \forall m$, are mutually uncorrelated. This leads to
\begin{equation}
p_{h_{m}\! | \text{h}} ( h_{m} | \text{h}; \bm{\theta}^{(i)})\! = \!\frac{ p_{\text{h} | h_{m}} \!(\text{h} | h_{m};\!\tilde{h}_m^{(i)}\!,\tilde{\sigma}_{\tilde{h}:m}^{2(i)}) \; p_{h_{m}}\!(h_{m}; \bm{\theta}^{(i)}) }{\! \int_{h'_{m}}\! p_{\text{h} | h_{m}} \!(\text{h} | h'_{m};\!\tilde{h}_m^{(i)},\tilde{\sigma}_{\tilde{h}:m}^{2(i)}) \; p_{h_{m}} \!(h'_{m};\! \bm{\theta}^{(i)}) }.
\label{eq:VGA_pdf_h_m_PDA}
\end{equation}

Next, by exploiting the assumption that $h_m$ follows a Bernoulli-Gaussian distribution and applying the Gaussian \ac{PDF} multiplication property \cite{Parker_TSP_2014}, equation \eqref{eq:VGA_pdf_h_m_PDA} can be equivalently expressed as \cite{VilaTSP2013}
\begin{equation}
p_{h_{m} | \text{h}} ( h_{m} | \text{h}; \bm{\theta}^{(i)})\! = \!(1\! -\! \hat{\rho}_{m}^{(i)}) \delta(h_m)\! +\! \hat{\rho}_{m}^{(i)}\mathcal{CN}(h_m;\hat{h}_{m}^{(i)},\hat{\sigma}_{h:m}^{2(i)}),
\label{eq:VGA_pdf_h_m_rewritten_BG_PDA}
\end{equation}
where
\begin{equation}
\vspace{-1ex}
\label{eq:BG_update_sparsity_rate}
\hat{\rho}_{m}^{(i)} \triangleq \Bigg( \frac{1\! -\! {\rho}^{(i)}}{{\rho}^{(i)}}  \frac{\tilde{\sigma}_{\tilde{h}:m}^{2(i)}\! +\! {\bar{\sigma}}^{(i)}}{\tilde{\sigma}_{\tilde{h}:m}^{2(i)}} \: e^{- \frac{|\tilde{h}_{m}^{(i)}|^2}{\tilde{\sigma}_{\tilde{h}:m}^{2(i)}} + \frac{|\tilde{h}_{m}^{(i)} - {\bar{h}}^{(i)}|^2}{\tilde{\sigma}_{\tilde{h}:m}^{2(i)} + {\bar{\sigma}}^{(i)}}} \!\!+ 1 \Bigg)^{\!\!-1}\!\!\!\!,
\end{equation}
with
\begin{equation}
\label{eq:BG_update_rules}
\hat{h}_{m}^{(i)} \triangleq \frac{{\bar{\sigma}}^{(i)} \tilde{h}_{m}^{(i)} + \tilde{\sigma}_{\tilde{h}:m}^{2(i)} {\bar{h}}^{(i)}}{\tilde{\sigma}_{\tilde{h}:m}^{2(i)} + {\bar{\sigma}}^{(i)}}\;\;\text{and}\;\;
\hat{\sigma}_{h:m}^{2(i)} \triangleq \frac{{\bar{\sigma}}^{(i)} \tilde{\sigma}_{\tilde{h}:m}^{2(i)}}{\tilde{\sigma}_{\tilde{h}:m}^{2(i)} + {\bar{\sigma}}^{(i)}}.
\end{equation}

Finally, in accordance with the prior specified for the channel coefficient $h_p$, we select, \ac{wlg}, a denoiser that omits the mean parameter, i.e., we set ${\bar{h}}^{(i)} = 0$ in equation \eqref{eq:h_m_pdf}, and, by extension, in equation \eqref{eq:BG_update_rules}\footnote{For completeness, the \ac{EM}-based parameter updates presented in the subsequent section will be derived for all parameters.}.
Based on \eqref{eq:VGA_pdf_h_m_rewritten_BG_PDA}, the soft replica $\hat{h}{m}^{(i)}$ and its associated \ac{MSE} $\hat{\sigma}{h:m}^{2(i)}$ can, in general, be computed from the conditional expectation as\footnote{Note that the damping mechanism has already been incorporated here, as also applied in Section \ref{secReceivercomm}, to mitigate convergence issues arising from erroneous hard-decision replicas.}
\begin{subequations}
\label{eq:soft_rep_and_MSE_updates}
\begin{equation}
\label{eq:PDA_soft_rep_update}
\hat{h}_m^{(i)} = \tilde{\beta}_h \hat{\rho}_{m}^{(i)} \hat{h}_{m}^{(i)} + (1 - \tilde{\beta}_h) \hat{h}_m^{(i-1)},
\end{equation}
\begin{equation}
\label{eq:PDA_MSE_update}
\hat{\sigma}_{h:m}^{2(i)} = \tilde{\beta}_h \big[(1 - \hat{\rho}_{m}^{(i)}) \hat{\rho}_{m}^{(i)} |\hat{h}_{m}^{(i)}|^2 + \hat{\rho}_{m}^{(i)} \hat{\sigma}_{h:m}^{2(i)}\big] + (1-\tilde{\beta}_h) \big[ \hat{\sigma}_{h:m}^{2(i-1)} \big].
\end{equation}
\end{subequations}

\subsubsection{Parameter Update via EM}

To iteratively refine the parameter set $\bm{\theta}$ of the Bernoulli-Gaussian distribution at each iteration, we employ the \ac{EM} algorithm.
Specifically, \ac{EM} is used as an iterative parameter estimation framework to determine the parameter vector $\bm{\theta}$ that maximizes the likelihood function $p_{\bar{\mathbf{r}} | \bm{\theta}} (\bar{\mathbf{r}}|\bm{\theta})$, where $\bar{\mathbf{r}}$ is defined in equation \eqref{eq:sensing_IO_1}, and $\bm{\theta}$ is specified in equation \eqref{eq:h_m_pdf}.

Following the approach in \cite{BishopBook2006}, we first express $p_{\bar{\mathbf{r}} | \bm{\theta}} (\bar{\mathbf{r}}|\bm{\theta})$ in a more tractable form by introducing latent variables, marginalizing with respect to an arbitrary \ac{PDF} $\hat{p}{\mathbf{{h}}} (\mathbf{h})$, and employing the log-likelihood function $\ln \big[ p{\bar{\mathbf{r}} | \bm{\theta}} (\bar{\mathbf{r}}|\bm{\theta}) \big]$. This formulation is then reorganized to emphasize the differential entropy and \ac{KL} divergence components, denoted as $\psi(\mathbf{h})$ and $\mathrm{D}\text{KL}\big(\hat{p}{\mathbf{{h}}} (\mathbf{h}) || p_{\mathbf{h} | \bar{\mathbf{r}} , \bm{\theta}} (\mathbf{h} | \bar{\mathbf{r}},\bm{\theta})\big)$, respectively, yielding
\begin{eqnarray}
\text{ln} \big[ p_{\bar{\mathbf{r}} | \bm{\theta}} (\bar{\mathbf{r}}|\bm{\theta})\big] \propto \int_\mathbf{h} \Big[ \hat{p}_{\mathbf{{h}}} (\mathbf{h}) \, \text{ln} \, p_{\mathbf{{h}},\bar{\mathbf{r}} , \bm{\theta}} (\mathbf{h},\bar{\mathbf{r}},\bm{\theta}) + \psi(\mathbf{h}) &&
\nonumber \\
&& \hspace{-33ex}  + \mathrm{D}_\text{KL}\big(\hat{p}_{\mathbf{{h}}} (\mathbf{h}) || p_{\mathbf{h} | \bar{\mathbf{r}} , \bm{\theta}} (\mathbf{h} | \bar{\mathbf{r}},\bm{\theta})\big) \Big].
\label{eq:log_likelihood_function_for_I}
\end{eqnarray}

Next, let us define the auxiliary function
\begin{equation}
\mathrm{J} \big( \hat{p}_{\mathbf{{h}}} (\mathbf{h}), \bm{\theta} \big) \triangleq \int_\mathbf{h} \Big[ \hat{p}_{\mathbf{{h}}} (\mathbf{h}) \, \text{ln} \, p_{\mathbf{{h}},\bar{\mathbf{r}} , \bm{\theta}} (\mathbf{h},\bar{\mathbf{r}},\bm{\theta}) + \psi(\mathbf{h}) \Big].
\label{eq:def_J_N_div}
\end{equation}

By leveraging the non-negativity property of the \ac{KL} divergence, we can derive the following lower bound for the aforementioned log-likelihood function
\begin{equation}
\text{ln} \big[ p_{\bar{\mathbf{r}} | \bm{\theta}} (\bar{\mathbf{r}}|\bm{\theta})\big] \geq \mathrm{J} \big( \hat{p}_{\mathbf{{h}}} (\mathbf{h}), \bm{\theta} \big).
\label{eq:non_neg_N_prop}
\end{equation}

The \ac{EM} algorithm summarized above allows us to maximize the log-likelihood function in a cost-effective manner by alternating between the E-step minimizing \ac{KL} divergence in equation \eqref{eq:log_likelihood_function_for_I} and the M-step maximizing the lower bound of the log-likelihood in equation \eqref{eq:non_neg_N_prop}, given by
\begin{subequations}
\begin{eqnarray}
\label{eq:E-step}
&&\hspace{-5ex}\text{E-step:}\quad \hat{p}_{\mathbf{{h}}} (\mathbf{h}) = \underset{\hat{p}'_{\mathbf{{h}}} (\mathbf{h})}{\mathrm{arg \; min}} \; J \big( \hat{p}'_{\mathbf{{h}}} (\mathbf{h}), \bm{\theta} \big),\\
\label{eq:M-step}
&&\hspace{-5ex}\text{M-step:}\quad \bm{\theta} = \underset{\bm{\theta}'}{\mathrm{arg \; max}} \; J \big( \hat{p}_{\mathbf{{h}}} (\mathbf{h}), \bm{\theta}' \big).
\end{eqnarray}
\end{subequations}

Concerning the E-step, we observe that solving equation \eqref{eq:E-step} is equivalent to minimizing over $\hat{p}_{\mathbf{{h}}} (\mathbf{h})$ for a fixed parameter vector $\bm{\theta}$. However, equation \eqref{eq:VGA_pdf_h_m_rewritten_BG_PDA} already provides the solution to this optimization problem.
Put differently, the Bernoulli-Gaussian distribution in equation \eqref{eq:VGA_pdf_h_m_rewritten_BG_PDA} constitutes the closed-form solution to the problem posed in \eqref{eq:E-step}.

Conversely, the M-step focuses on optimizing $\bm{\theta}$ given a fixed distribution $\hat{p}_{\mathbf{{h}}} (\mathbf{h})$. By leveraging \eqref{eq:VGA_pdf_h_m_rewritten_BG_PDA}, the maximization task in \eqref{eq:M-step} can be equivalently reformulated as
\begin{equation}
\bm{\theta} = \underset{\bm{\theta}'}{\mathrm{arg \; max}} \; \mathbb{E}_{\mathbf{h} | \text{h}} \Big\{ \text{ln} \big[ p_{\mathbf{{h}},\bar{\mathbf{r}} , \bm{\theta}} (\mathbf{h},\bar{\mathbf{r}},\bm{\theta}') \big] |\; \text{h} ; \bm{\theta}^{(i)}  \Big\},
\label{eq:reform_M-step}
\end{equation}
and solved efficiently via a Lagrange method, yielding the update rules
\begin{subequations}
\begin{equation}
{\rho}^{(i)} = \frac{1}{K_\tau D_\nu} \sum_{m=1}^{K_\tau D_\nu} \hat{\rho}_{m}^{(i)},
\label{eq:Update_for_lambda}
\vspace{-0.75ex}
\end{equation}
\begin{equation}
{\bar{h}}^{(i)} = \frac{1}{{K_\tau D_\nu} \cdot {\rho}^{(i)}} \sum_{m=1}^{K_\tau D_\nu} \hat{\rho}_{m}^{(i)} \hat{h}_{m}^{(i)},
\label{eq:Update_for_mu}
\vspace{-0.75ex}
\end{equation}
\begin{equation}
{\bar{\sigma}}^{(i)} = \frac{1}{{K_\tau D_\nu} \cdot {\rho}^{(i)}} \sum_{m=1}^{K_\tau D_\nu} \hat{\rho}_{m}^{(i)} \bigg(  \big|  \hat{h}_{m}^{(i)} - {\bar{h}}^{(i)}  \big|^2 + \hat{\sigma}_{h:m}^{2(i)} \bigg).
\label{eq:Update_for_phi}
\vspace{-0.5ex}
\end{equation}
\label{eq:Complete_set_update_BG}
\end{subequations}

\begin{algorithm}[t]
\caption{PDA-based Radar Parameter Estimation (Prop.)}
\label{alg:PDA-EM}
\textbf{Input:} Receive signal $\bar{\mathbf{r}}$, dictionary matrix $\mathbf{E}$, noise power $N_0$, number of paths $P$, average channel power per path $\sigma_h^2$, maximum number of iterations $i_\text{max}$ and damping factor $\tilde{\beta}_h$. \\
\textbf{Output:} Estimates $\hat{\tau}_p$ and $\hat{\nu}_p$ extracted from the non-zero indices of the sparse channel estimate vector $\!\hat{\;\mathbf{h}}$.\\
\textbf{Initialization:} Set counter to $i = 0$, set initial distribution parameters ${\rho}^{(0)} = P/(K_\tau D_\nu)$ and ${\bar{\sigma}}^{(0)} = 1/P$, set average channel power per path $\sigma_h^2 = 1/(K_\tau D_\nu)$, and set initial estimates $\hat{h}_{m}^{(0)} = 0$ and ${\hat{\sigma}}_{h:m}^{(0)} = \sigma_h^2, \forall m$.
\vspace{0.4ex}
\hrule

\begin{algorithmic}[1]

\STATEx \hspace{-3.8ex}{\textbf{for}} {$i=1$ to $i_\text{max}$} {\textbf{do}} $\forall m$
\STATE Compute soft signal vectors $\tilde{\mathbf{r}}_{h:m}^{(i)}$ from equation \eqref{eq:Soft_IC_PDA}.
\STATE Compute soft extrinsic channel beliefs $\tilde{h}_m^{(i)}$ and their
variances $\tilde{\sigma}_{\tilde{h}:m}^{2(i)}$ from equation \eqref{eq:mean_and_var_extrinsic_belief_PDA}.
\STATE Compute denoised sparsity rates $\hat{\rho}_m^{(i)}$ from eq. \eqref{eq:BG_update_sparsity_rate}.
\STATE Compute denoised channel estimates $\hat{h}_{m}^{(i)}$  and their variances $\hat{\sigma}_{h:m}^{2(i)}$ from equation \eqref{eq:BG_update_rules}.
\STATE Compute damped channel estimates $\hat{h}_{m}^{(i)}$ and variances $\hat{\sigma}_{h:m}^{2(i)}$ from equation \eqref{eq:soft_rep_and_MSE_updates}.
\STATE Update distribution parameters ${\rho}^{(i)}$ and ${\bar{\sigma}}^{(i)}$ from eq. \eqref{eq:Complete_set_update_BG}.
\STATEx \hspace{-3.8ex}{\textbf{end}} {\textbf{for}}
\end{algorithmic}
\label{PDA-EM}
\noindent Compute the estimates $\hat{\tau}_p$ and $\hat{\nu}_p$ corresponding to the indices $m$ of the non-zero entries of $\!\hat{\;\mathbf{h}}$ in accordance to equation \eqref{eq:sparse_channel_vector}.
\end{algorithm}

A complete and compact description of the proposed \ac{PDA}-based \ac{RPE} scheme is given in the form of a pseudocode in Algorithm \ref{alg:PDA-EM} on the following page.
Upon convergence of the above procedure, the estimate of the sparse vector $\mathbf{h}$ is known. Finally, the $(i,j)$ indices of $\mathbf{h}$ can be used to obtain the corresponding delay and Doppler shifts, and hence the ranges and velocities of the objects in the vicinity.

\subsection{Complexity Analysis}
\label{sec:RPE_Complexity}

The computational complexity of the proposed \ac{PDA}-based \ac{RPE} method is on the order of $\mathcal{O}(N^3)$.
An important observation is that this complexity is essentially independent of the delay-Doppler grid size determined by $K_\tau$ and $D_\nu$. 
Consequently, the method can accommodate significantly larger grids, even when multiple iterations are employed as in \cite{Ranasinghe_ICASSP_2024}.
This is because the dominant operation is a matrix inversion whose cost does not scale with $K_\tau \cdot D_\nu$, thereby allowing high-resolution parameter estimation to be performed efficiently.

\subsection{Performance Analysis}

\begin{figure}[H]
\includegraphics[width=\columnwidth]{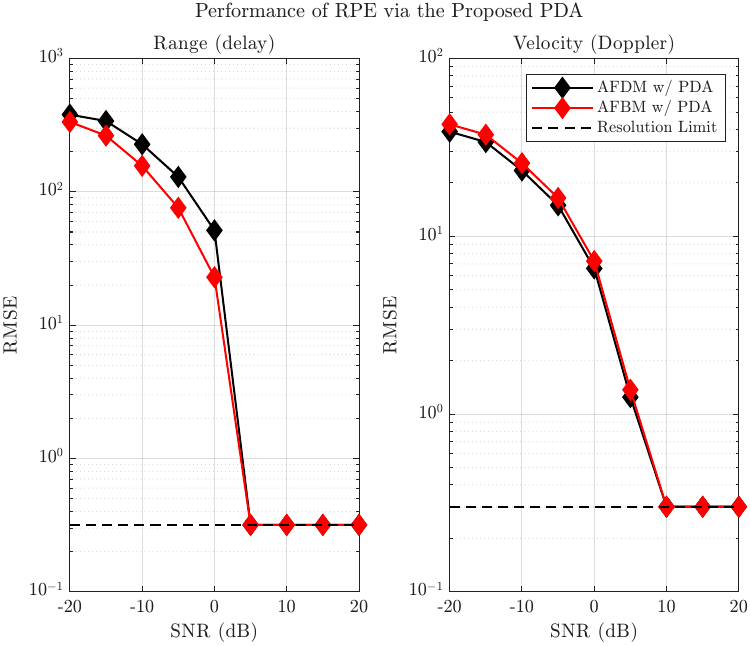}%
\vspace{-2ex}
\caption{Radar parameter estimation performance of the proposed \ac{AFBM} waveform compared to \ac{AFDM}, in terms of \ac{RMSE} for target range and velocity with $R=3$, via the proposed \ac{PDA}-based approach with $P = 256$.}
\label{fig:ISAC_PDA}
\end{figure}

Finally, Fig.~\ref{fig:ISAC_PDA} presents the \ac{RMSE} performance of the proposed \ac{PDA}-based estimator when applied to both \ac{AFDM} and \ac{AFBM} waveforms. 
The results indicate that the two schemes achieve comparable accuracy in estimating both range and velocity parameters. 
A slight improvement in range estimation is observed with \ac{AFBM}, which is consistent with the lower sidelobes in ambiguity function analysis presented earlier.

\section{Conclusion}
\label{secConclusion}

We introduced an affine-domain filter-bank scheme, termed \ac{AFBM}, is introduced. 
Building on this waveform, novel receiver designs for both communications and sensing, collectively referred to as \ac{ISAC}, were developed. 
For the communications task, a low-complexity \ac{GaBP}-based data detection algorithm was proposed, requiring only element-wise scalar operations. 
For the sensing task, an \ac{RPE} algorithm was employed, leveraging an \ac{EM}-assisted \ac{PDA} framework to accurately estimate target ranges and velocities. 
Through analytical and numerical evaluations, the proposed \ac{AFBM} scheme was shown to achieve advantages in terms of \ac{PAPR}, \ac{OOBE}, \ac{AF}, \ac{BER}, and \ac{RMSE}, thereby demonstrating clear improvements over the conventional \ac{AFDM} waveform. 
Future research directions include the development of adaptive channel estimation schemes, as well as investigations into the robustness of \ac{AFBM} under high-mobility conditions. 
Additionally, the design of native chirp-parameter optimization strategies will be explored to further reduce both \ac{PAPR} and \ac{OOBE}. 
Overall, the results highlight \ac{AFBM} as a promising candidate waveform for \ac{6G} \ac{ISAC} systems, combining communication reliability, sensing accuracy, and spectral efficiency in a unified framework.

\balance
\selectlanguage{english}
\bibliographystyle{IEEEtran}
\bibliography{references.bib}

\end{document}